\documentclass[11pt,preprint]{revtex4}
\usepackage{graphicx}% Include figure files
\usepackage{dcolumn}% Align table columns on decimal point
\usepackage{bm}% bold math

\begin{document}

\title{Modification of single molecule fluorescence close to a nanostructure: radiation pattern, spontaneous emission and quenching}

\author{S. K\"{u}hn}
\altaffiliation{Present address: University of California at Santa
Cruz, CA 95067 USA}

\author{G. Mori}
\affiliation{Laboratorium f\"{u}r Physikalische Chemie, ETH
Z\"{u}rich, 8093 Z\"{u}rich, Switzerland}

\author{M. Agio}
\affiliation{Laboratorium f\"{u}r Physikalische Chemie, ETH
Z\"{u}rich, 8093 Z\"{u}rich, Switzerland}

\author{V. Sandoghdar}
\email{vahid.sandoghdar@ethz.ch} \affiliation{Laboratorium f\"{u}r
Physikalische Chemie, ETH Z\"{u}rich, 8093 Z\"{u}rich, Switzerland}

\begin{abstract}
\boldmath The coupling of nanostructures with emitters opens ways
for the realization of man-made subwavelength light emitting
elements. In this article, we investigate the modification of
fluorescence when an emitter is placed close to a nanostructure. In
order to control the wealth of parameters that contribute to this
process, we have combined scanning probe technology with single
molecule microscopy and spectroscopy. We discuss the enhancement and
reduction of molecular excitation and emission rates in the presence
of a dielectric or metallic nanoparticle and emphasize the role of
plasmon resonances in the latter. Furthermore, we examine the
spectral and angular emission characteristics of the
molecule-particle system. Our experimental findings are in excellent
semi-quantitative agreement with the outcome of theoretical
calculations. We express our results in the framework of optical
nanoantennae and propose arrangements that could lead to the
modification of spontaneous emission by more than 1000 times.
\end{abstract}

\maketitle

\bigskip

\section{Introduction}

The radiative properties of an emitter, such as the angular
distribution of its emitted power, its spectrum and fluorescence
lifetime can be strongly modified in confined geometries. One of the
earliest indications of these phenomena was pointed out by A.
Sommerfeld who considered the radiation of a dipole close to a
surface~\cite{Sommerfeld1909}. Another seminal proposal was made by
E. M. Purcell on the enhancement of the decay rate of an atomic
excited state inside a cavity~\cite{Purcell:46}. The pioneering
experimental work dates back to around 1970 when K. H. Drexhage
showed that the fluorescence lifetimes of emitters placed very close
to a flat mirror were different from those in free space
\cite{Drexhage:74a}. In the 1980s and 1990s several groups have
demonstrated the possibility of controlling radiative decay rates
and the emission pattern by putting emitters in confined geometries
such as the spaces between two flat substrates, between the mirrors
of high-finesse cavities and in whispering gallery mode
resonators~\cite{Berman, Chang03}. In all these cases the dimensions
of the geometries surrounding the emitter have been superior to the
transition wavelength of interest. Considering that advances in
nanosciences over the past decades have made the understanding and
manipulation of the nanometer scale processes both more accessible
and desirable, it is highly interesting to investigate the radiative
properties of atoms, ions, molecules or quantum dots using
subwavelength boundaries. At our laboratories, we have conducted
both theoretical~\cite{henkel:98, Rogobete:03, Koenderink:05b,
Rogobete:07} and experimental~\cite{Schniepp:02, Kuehn:06,
Kuehn:06b, Seelig2007} research in this direction. Before we present
some of our most recent results, we find it instructive to provide a
brief historical overview of some of the relevant activities in the
past three decades.

A few years after the work of Drexhage on the change of the
fluorescence lifetime, researchers found out experimentally that the
the Raman scattering signal from molecules can be enhanced close to
rough metallic surfaces~\cite{Fleischmann1974}. Soon after this
discovery, it was proposed that an important contribution to the
observed effect can come from the enhancement of the electromagnetic
fields close to sharp edges of a rough surface~\cite{Jackson-book}.
Furthermore, it was pointed out that this lightning rod type of
effect can be strengthened if surface plasmon resonances are
supported by the system~\cite{Metiu:84,Moskovits:85}. In order to
model the situation at hand, scientists simplified a rough surface
by considering single or small aggregates of metallic nanoparticles
such as spheres and ellipsoids. It was also pointed out that not
only the excitation channel is enhanced due to the strong near
fields, but also the emission of a dipole is modified close to a
nanoparticle~\cite{Gersten1981, Ruppin1982, chew:88}, much like the
effect observed by Drexhage. Early experimental efforts of the 1970s
and 1980s used ensembles of molecules and nanoparticles, but the
inherent inhomogeneities in such experiments plagued clear
comparisons with the theoretical predictions. The main obstacle is
that the radiative properties of an emitter such as a molecule
depend very strongly on its orientation and separation from the
nanostructure as well as the shape, material and size of the latter.
In addition, the wavelength dependence of the dielectric constants
of the nanostructure determines the strength of the modifications at
the emission and absorption wavelengths of the emitter.

Advances in nano-optics have brought about various new tools for
studying single molecules~\cite{Kulzer:04} and single metallic
nanoparticles~\cite{klar:98, Kalkbrenner:04, Lindfors:04,
vanDijk:06} both spatially and spectrally. These developments
motivated the investigation of the Raman enhancement at the level of
a single molecule and a single nanoparticle~\cite{Nie:97} or small
aggregates~\cite{kneipp:97}. In particular, scanning probe
techniques offered a convenient way of controlling the coupling
between a molecule and a nanostrcuture. The first impulse of this
method was to modify the fluorescence lifetime of a fluorescent
molecule by approaching it to a metallized near-field
tip~\cite{Ambrose:94, xie:94}. At about the same time sharp extended
tips were introduced as subwavelength sources of field enhancement
for obtaining better resolution in near-field
microscopy~\cite{Kawata1994, Zenhausern1994, gleyzes:95}. Later, the
same arrangement was successfully applied to subwavelength studies
of SERS on single nano-objects such as carbon
nanotubes~\cite{Hartschuh:03}. However, experience has shown that
production of well-defined and reproducible tips is a very tedious
task~\cite{Sanchez:99}. Furthermore, a theoretical treatment of
extended tips is nontrivial and could lead to numerical artefacts.
To remedy these problems, several years ago we set out to perform
experiments where a well characterized metallic nanoparticle is
attached to a dielectric tip~\cite{Kalkbrenner:01} to act as a
´´local enhancer´´ that can be positioned at will in front of a
single molecule~\cite{Kuehn:06}, as originally proposed by J. Wessel
in 1985 in the context of SERS and nonlinear near-field
microscopy~\cite{wessel:85}. As we show in what follows, this
strategy has allowed us to control the relative distances and
orientations of single molecule and gold nanosphere.

In the past few years, many of the ideas mentioned above have been
reformulated in a new framework based on the concepts from antenna
theory~\cite{Pohl:00}. Different groups have worked to extend the
conventional antenna designs to the nanometer scale~\cite{Schuck:05,
Muhlschlegel2005} although the very different behavior of metals at
optical and microwave or radio frequencies makes this extension
nontrivial. It turns out that again a single gold nanoparticle that
supports plasmon resonances can act as a very simple and elementary
antenna. In this article we show that such a subwavelength particle
can serve as a receiving and transmitting subwavelength resonant
antenna with a dipolar radiation pattern and a well-defined
resonance spectrum. The great advantage of this system is that its
scattering properties and plasmon spectra are well described
analytically using Mie theory~\cite{Kreibig,Kalkbrenner:04}. In the
last section of this article, we discuss realistic designs for
two-particle antennae and how one could reduce the effect of
quenching due to losses in metals.

The paper is organized as follows. We will first introduce a simple
model of a dipole close to a sphere as the basis for our
calculations and the interpretation of the experimental results. In
the following section we will present a selection of scanning probe
near-field images of single molecules. We will start with a simple
glass tip and establish the modification of excitation and emission
properties in the absence of material absorption. Then we will focus
specifically on the quenching phenomenon by adding a thin layer of
chromium onto the tip. The choice of chromium deliberately avoids
resonances. We then move on to using a tip that carries a single
gold nanoparticle at its apex. This tip produces the strongest
effect on the excitation, the decay rate, the emission spectrum and
the radiation pattern of a molecule reported to date. To complete
these studies we will present design guidelines and experimental
results on resonant antenna structures involving two nanoparticles.

\section{Theoretical concepts}

\begin{figure} \centering
\includegraphics[width = 5 cm]{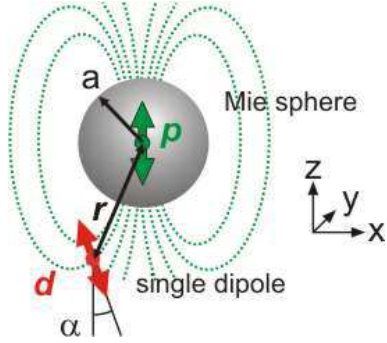}
\caption{A nanosphere of radius $a$ placed close to a dipolar
emitter. $r$ is distance between the sphere center and the emitter,
and $\alpha$ denotes the tilt angle of the emitter's dipole moment.}
\label{np-geom}
\end{figure}

In this section, we briefly discuss the influence of a nanostructure
on the interaction between an incident laser field and a fluorescent
molecule, which we treat as a classical oscillating dipole (see
fig.~\ref{np-geom}). Generally, the fluorescence signal $S_f$ from
the molecule can be expressed as \cite{Thomas2004}:

\begin{equation}
S_f({\bf \hat d, r}) = c \; \xi({\bf \hat d, r}) \, K({\bf \hat d, r}) \, \eta({\bf \hat d, r})  \label{fl-enh}
\end{equation}
for the weak excitation regime far from saturation. Here the
coefficient $c =  S_0 / (\xi_0 \eta_0)$ normalizes the signal to the
fluorescence $S_0$ of the unperturbed molecule, $\xi$ is the
collection efficiency which can also depend on the system
parameters, and $K$ is the ratio of the excitation rates in presence
and absence of the tip given by
\begin{equation}
K({\bf \hat d, r}) = |{\bf d} \cdot {\bf E_{loc}}({\bf r})|^2/|{\bf
d} \cdot {\bf E_{inc}}({\bf r})|^2~. \label{exc-enh}
\end{equation}
Finally, $\eta$ is the apparent quantum yield, defined as
\begin{equation}
\eta({\bf \hat d, r}) = \frac{\gamma_{rad}({\bf \hat d, r})}{\gamma_{rad}({\bf \hat d, r}) + \gamma_{nr}({\bf \hat d, r})}
\label{em-enh}
\end{equation}
where $\gamma_{rad}$ and $\gamma_{nr}$ denote the radiative and
nonradiative decay rates of the excited state lifetime. Thus, the
key parameters are the field ${\bf E_{loc}}({\bf r})$ around the tip
and the decay rates $\gamma_{rad}({\bf d, r})$ and $\gamma_{nr}({\bf
d, r})$ in the presence of the nanostructure.

In the case of a spherical or ellipsoidal particle, the
nanostructure can be fully described using generalized Mie theory
for different incident fields~\cite{Bohren1998, Tai1971,
Zvyagin1998, Quinten1999, Lock1995}. Further simplifications apply
for small spheres if $k a \ll 1$ where $k$ is the magnitude of the
relevant wavevector and $a$ is the radius of the sphere. The optical
response is then identical to that of a dipole with the moment ${\bf
p} = 4 \pi \alpha {\bf E}_{inc}$ where $\alpha$ is the electrostatic
polarizability of the sphere and $E_{inc}$ denotes the incident
field~\cite{Bohren1998, Kelley2003}. The resulting field around the
sphere is then given by ${\bf E_{loc}}({\bf r}) = {\bf E}_s({\bf p,
r})+{\bf E}_{inc}({\bf r})$ where $E_s$ is the scattered field due
to $\bf p$. Metallic nanoparticles can respond resonantly upon the
excitation of surface plasmon-polaritons, which are the optical
pendant to the resonances of radio antennae~\cite{Bohren1998}. As we
will see below, the interaction of various dielectric and metallic
tips can be described to a good degree by approximating them as
small spheres.

Using this model in conjunction with eqn.~\ref{exc-enh} has the
following consequences. The magnitude of the excitation enhancement
depends on the position $\bf r$ and the relative orientations
between $\bf p$ and the molecular dipole moment $\bf d$. The highest
excitation enhancement is expected when $\bf {\hat r}$, $\bf {\hat
p}$ and $\bf {\hat d}$ are aligned. In the quasi-static dipole
approximation, the magnitude of $K$ then goes as $1/r^6$ and reaches
a finite maximum on the surface (i.e. $r = a$), independent of the
size of the sphere. Close to the surface, $E_s$ will dominate over
$E_{inc}$ provided that there is an appreciable polarization. At
distances where the magnitudes of $E_s$ and $E_{inc}$ are similar
their interference leads to a modulation of $K$~\cite{Richter1999}.

The emission of the molecule is also altered in the presence of the
sphere ~\cite{Chew1979,Lakowicz2005} via a change in the radiative
decay rate $\gamma_{rad}$~\cite{Bian1995, Sullivan1997, Metiu1984,
Klimov1996} and in the non-radiative rate $\gamma_{nr}$ if the
material of the sphere absorbs light~\cite{Chew1987}. Without
engaging in a detailed discussion, we make the following remarks
that can provide a quick intuitive picture of the physics at hand.
The presence of a nanoparticle can result in a change of the
apparent quantum yield $\eta$ (eqn.~\ref{em-enh}) of the joint
molecule-sphere system~\cite{Das2002, Metiu1984, Agio2007}. $\eta$
may be increased for a molecule with a small initial yield $\eta^0 <
1$~\cite{Agio2007} but can only decrease if $\eta^0 \approx 1$.
Furthermore, $\gamma_{nr}({\bf d, r})$ will increase as $1/(r-a)^6$
when the molecule approaches the absorbing surface, leading to the
notorious fluorescence quenching for $r \rightarrow a$. The emission
and the excitation processes are linked in accordance with the
reciprocity theorem~\cite{Caticha2000, Carminati1998}. In other
words, an antenna effect in the enhancement of the excitation also
leads to the enhancement of the emission rate. Again, maximal
radiation is achieved when ${\bf \hat d}$ is oriented radially and
is parallel to $\bf {\hat p}$~\footnote{Note that field $\bf
E_{inc}$ now originates from the emitting molecule}. ${\bf d}$ and
${\bf p}$ then add up to form a large dipole moment. The opposite is
the case if ${\bf \hat d}$ lies tangential to the sphere surface.
Here the resulting $\bf \hat p$ becomes antiparallel to ${\bf \hat
d}$ and their summation is destructive. The magnitude of $\bf p$ has
a $1/r^6$ dependence to the lowest (dipole-) approximation.
Furthermore, the spectral dependence of $\epsilon(\lambda)$ and thus
of the particle plasmon resonance also dictate a spectral variation
of $\gamma_{rad}(\lambda)$ and $\gamma_{nr}(\lambda)$ . The
different reference points $r = 0$ for $\gamma_{rad}$ and $K$, and
$r = a$ for $\gamma_{nr}$ lead to an intersection of the distance
dependence curves of the enhancement and dissipation values. As a
result, there exists a maximum of $S_f$ at a finite distance from
the surface whereas $\gamma_{tot}$ will increase monotonously.

\section{Experimental}

\begin{figure} \centering
\includegraphics[width = 8 cm]{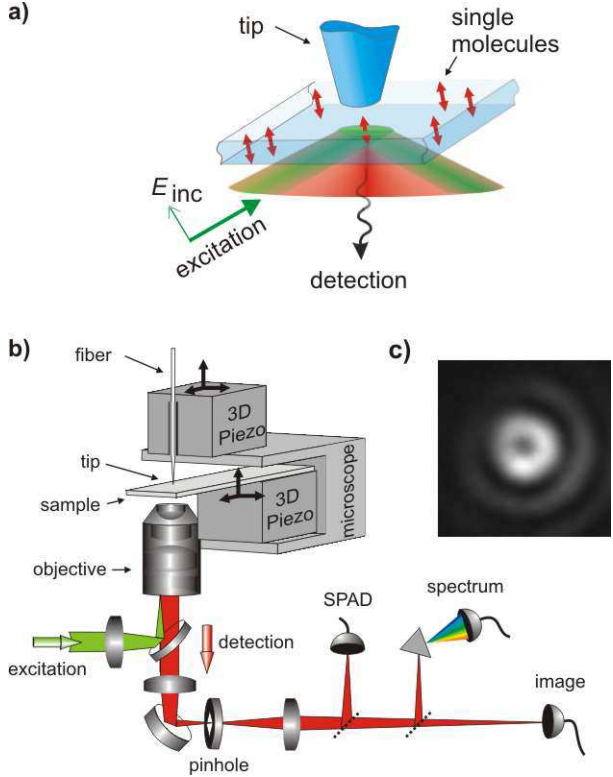}
\caption{a) Schematic arrangement of the experiments discussed in
this article. A tip is scanned across a thin sample containing
single molecule. b) The experimental setup consisting of the
scanning stages for the tip and the sample. c) Far-field
fluorescence image of a single terrylene molecule. The doughnut
emission pattern is a signature of the nearly vertical orientation
of the terrylene molecules in the sample.} \label{setup}
\end{figure}

The experimental setup is described in detail in
Refs.~\cite{Kalkbrenner:01} and~\cite{Kuehn:06}. It was designed to
provide full control over the relative displacement $\bf r$ between
a single molecule placed on a substrate and the extremity of a tip.
We used a home-built scanning probe microscope based on shear-force
distance control~\cite{Betzig1992, Karrai2000} mounted on an
inverted optical microscope. Single-molecule samples were prepared
from a dilute solution of terrylene and para-terphenyl (pT) in
toluene that was subsequently spin coated onto a microscope slide
according to the procedure described in Ref.~\cite{Pfab:04}. The
solution crystallizes into a film that can be as thin as about $15$
nm in height with well dispersed terrylene molecules that assume a
polar angle of $\alpha = (15 \pm 5)^\circ$~\cite{Buchler:05,
Kummer1997} along the $c$-axis of pT (see fig.~\ref{setup}).
Terrylene turns out to be remarkably photostable in this sample,
allowing long term and repeated measurements on the same single
molecule~\cite{Pfab:04}. Furthemore, terrylene has a quantum yield
near unity~\cite{Buchler:05}.

For illumination and collection we used an immersion objective
(Zeiss, NA 1.4) in different quickly switchable modes. We
investigated the plasmon resonances of gold nanoparticles using dark
field illumination from a Xe arc-lamp and a spectrometer with a
cooled CCD detector. Excitation was typically achieved in total
internal reflection mode which produced an axial polarization in the
focal plane for p-polarized light. For lifetime measurements by time
correlated single photon counting~\cite{Becker2004} we used a single
photon avalanche diode (SPAD) and laser pulses of $< 30$ ps duration
at repetition rates of $5-75$ MHz. Alternatively we could switch to
a cooled imaging CCD camera to record the spatial emission pattern
at a system magnification of 230~\cite{Bohmer2004}.

Typical experiments started with the imaging of single molecules in wide-field mode to establish their orientation and
signal strength $S_0$. A single molecule was then selected by closing a pinhole of variable size located in an
intermediate image plane. Eventually, a tip was approached and scanned over the stationary sample at distances down to $2$
nm from the surface. At each scan pixel, fluorescence intensity and lifetime were recorded simultaneously.

\section{Results and Discussion}

\subsection{Single molecules and dielectric tips}
The simplest treatments of the modification of spontaneous emission
consider perfect metallic mirrors~\cite{Haroche-Houches}, but since
real metals absorb in the visible domain, the fluorescence behavior
of an emitter is also affected by quenching. In order to avoid this
complication, one can use dielectric mirrors~\cite{Sullivan1997} or
nanostructures~\cite{henkel:98}. Indeed, a number of groups have
studied the effect of various dielectric tips on single emitters.
While some researches~\cite{Trabesinger:03b, Trabesinger:02} have
reported quenching of fluorescence, others have observed
enhancement~\cite{Protasenko2004, H'dhili2002}. In this section, we
report on well-controlled experiments on aligned single molecules
and a glass tip. We show that a simple theoretical model provides a
satisfactory agreement with our measurements and indicates the
importance of the orientation of the molecule in the modification of
its fluorescence under a tip. Details of these investigations can be
found in Ref.~\cite{Kuehn:06b}.

\begin{figure}\centering
\includegraphics[width= 8 cm]{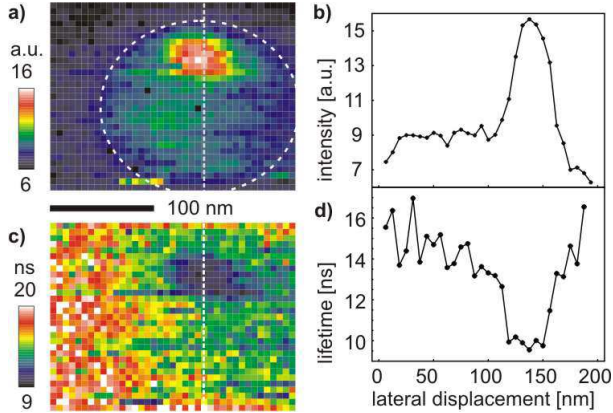} \caption{a) Flourescence near-field image of a single molecue using a glass
tip. b) Cross-section as indicated. c) Map of the lifetime
corresponding to image a). d) Cross-section as indicated.}
\label{glass-scan}
\end{figure}

A bare heat-pulled glass tip was approached and scanned over a
single terrylene molecule embedded in a thin pT film in constant gap
mode. The resulting map of the fluorescence signal $S_f$ is shown in
fig.~\ref{glass-scan}a). We can identify two regions where the tip
causes $S_f$ to increase: an area of $150 \times 150$ nm$^2$ with
moderate enhancement indicated by the white circle and then a sharp
increase by a factor of 2 in $S_f$ over $35 \times 35$ nm$^2$ in the
upper part of this circle. The presence of these two distinct
regions can be attributed to the tip morphology. Tips created by the
heat-pulling process can exhibit plateaus of $50 - 300$ nm in size
depending on the pulling parameters and sometimes carry spikes of
several 10 nm. Both plateau and spike act as polarizable
nano-objects and thus increase the excitation rate of the molecule
by $K$. In the absence of absorption and far below saturation this
translates into a proportional increase of $S_f$ according to
eqn.~\ref{fl-enh}. We note that the high collection efficiency
$\xi_0 \approx 75\%$ implies that $\xi({\bf d, r})$ does not play a
significant role in the observed enhancement~\cite{Kuehn:06}. The
different magnitudes of $K$ can be partly attributed to a difference
in shape and partly to the gap by which the plateau of the tip is
separated from the molecule due to the presence of asperity. Sharp
features generally give rise to stronger enhancements than dull
ones~\cite{H'dhili2002} due to a stronger lightning-rod effect.

\begin{figure} \centering
\includegraphics[width= 6 cm]{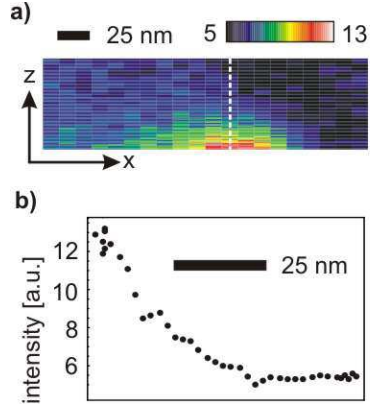}
\caption{a) A $xz$ section of the enhancement landscape taken by
approaching the tip to a single molecule. b) Intensity trace when
the tip is approached in $z$ as indicated in a).} \label{glass-appr}
\end{figure}

The dependence on the vertical separation between the tip and the
molecule is displayed in fig.~\ref{glass-appr}a) where a glass tip
is repeatedly approached to a terrylene molecule in $\bf \hat z$ at
different lateral separations $y$. As seen from a cross section of
this data in fig.~\ref{glass-appr}b), the enhancement is only
significant within the last $15$ nm from contact. This fact provides
a direct evidence for the strong distance dependence of the near
fields close to a polarizable object and supports our interpretation
of the lateral scan image.

During the scan we also recorded the excited state lifetime $\tau =
\gamma_{tot}^{-1}$ (see fig.~\ref{glass-scan}c)). The fluorescence
lifetime drops from $18$ ns for a single molecule embedded in a thin
film to 16 ns in the larger area and down to 9 ns in the upper part.
Given that glass has a negligible absorption at the molecular
emission wavelength, the shorter lifetime is a direct evidence for
the accelerated radiation and can be interpreted as an antenna
effect~\cite{Martin2001, Patane2004}. We emphasize, however, that
below saturation this increase does not cause a change in $S_f$,
which is limited by the excitation rate. Thus, the simultaneously
observed increase of fluorescence in fig.~\ref{glass-scan}a) is due
to the enhancement of $K$.

\begin{figure} \centering
\includegraphics[width= 8 cm]{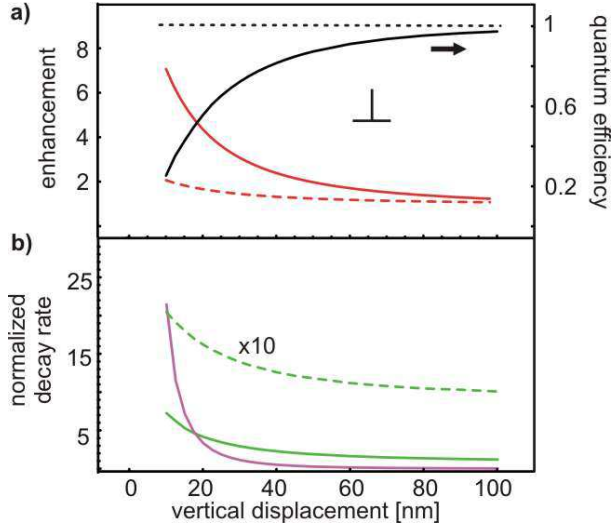}
\caption{Emission properties of a dipole placed close to a sphere at radial orientation. Dipole emission wavelength: $580$
nm, sphere diameter: $100$ nm, material: chromium - \textit{solid lines}, glass - \textit{broken lines}. a) Fluorescence
enhancement (\textit{red lines}) and quantum yield (\textit{black lines}). b) Radiative (\textit{green}) and non-radiative
(\textit{pink}) decay rate normalized to the free dipole.} \label{tip-theo-1}
\end{figure}

The magnitudes of the enhancement and lifetime change are well
reproduced in a model where we approximate the tip by a glass sphere
that has a diameter of $100$ nm and a refractive index of 1.5. In
this simple treatment, we disregard the interface of pT and air and
let the system be illuminated by a plane wave along ${\bf \hat x}$.
For the emission we choose the main peak in the terrylene spectrum
at $580$ nm. Using the formalism described above, we have evaluated
the excitation enhancement $K$, the decay rates $\gamma_{rad/nr}$
and the quantum yield $\eta$ as a function of the vertical
molecule-tip separation $z$~\cite{Chew1987}. The broken lines in
figs.~\ref{tip-theo-1}a) and b) show the results for a molecule
oriented perpendicular to the surface. In the absence of absorption
$\eta$ (black line) remains constant throughout. $K$ and
$\gamma_{rad}$ (red and green curves) increase by up to a factor of
2 at a distance $z = 10$ nm. The lateral extent of the enhancement
roughly corresponds to the radius of the particle. It is quite
remarkable that this simple model predicts the experimentally
measured values quite accurately.

\begin{figure} \centering
\includegraphics[width= 8 cm]{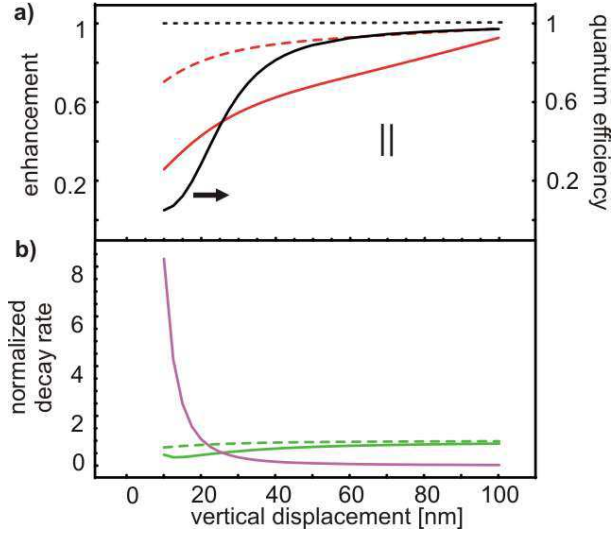}
\caption{Emission properties of a dipole placed close to a sphere at tangential orientation. parameters as in
fig.~\ref{tip-theo-1}, material: chromium - \textit{solid lines}, glass - \textit{broken lines}. a) Fluorescence
enhancement (\textit{red lines}) and quantum yield (\textit{black lines}). b) Radiative (\textit{green}) and non-radiative
(\textit{pink}) normalized decay rate.} \label{tip-theo-2}
\end{figure}

We also applied the formalism to a tangentially oriented molecule. The results are shown in figs.~\ref{tip-theo-2} a) and
b), again as broken lines. Here the model predicts a mild decrease in the excitation rate which means suppression of $S_f$
to about $70\%$. Thus, it is the orientation of the molecule that decides if a near-field image with a dielectric tip will
show enhancement or inhibition. The difference for the two orientations can easily be illustrated in a simple picture
where the molecular dipole interacts with its image dipole. The situation changes, however, if the tip material starts to
absorb~\cite{Chance1978} as is the case for metals.

\subsection{Single molecules and metallic extended tips}

Metallic structures are particularly attractive for modifying fluorescence properties of emitters because they have large
dielectric constants $|\epsilon(\omega)|$, leading to a strong lightning-rod effect~\cite{Ermushev1993}. For chromium,
$\epsilon \approx -8.5 + 29i$~\cite{Lide1995} leads to an increase by a factor of 20 compared to glass in the limit of an
infinitely thin needle~\cite{Klimov2001}. However, this gain in excitation can be overshadowed by quenching caused by the
imaginary part of $\epsilon$~\cite{Das2002, Azoulay2000}. The shape of the structure, the orientation of the molecule and
their separations determine the final fluorescence modification.

\begin{figure}\centering
\includegraphics[width= 8 cm]{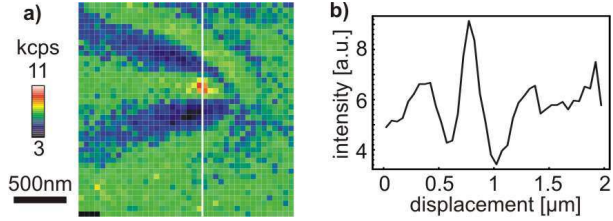}
\caption{Single molecule fluorescence near-field images in
constant-gap mode for a chromium coated tip. Molecule oriented
normal to the scanning plane (terrylene in para-terphenyl).}
\label{cr-scan-1}
\end{figure}

Metallic tips  have been used in many near-field
experiments~\cite{inouye:94,Sanchez:99,Kramer2002}. In our work, we
metalized heat-pulled glass tips by head-on thermal evaporation of
$30-40$~nm of chromium. Figure~\ref{cr-scan-1}a) shows an image of
$S_f$ in the same fashion as for the glass tip. A cross section of
this image is provided in fig.~\ref{cr-scan-1}b) for a quantitative
scrutiny. Two features can be distinguished: a sharp peak of $150$
nm in full width at half maximum and a far-field interference
pattern. The fluorescence enhancement amounts to only $1.6$.

\begin{figure}\centering
\includegraphics[width=8 cm]{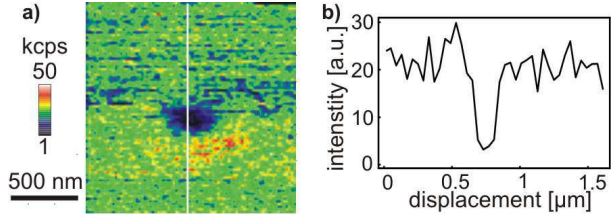}
\caption{Single molecule fluorescence near-field images in
constant-gap mode for a chromium coated tip. Molecule oriented in
the scanning plane (DiI in PMMA).} \label{cr-scan-2}
\end{figure}

Next we exchanged the sample with a thin film of
PMMA~\footnote{PMMA: polymethyl methacrylate} doped with randomly
oriented DiI molecules~\footnote{DiI: dioctadecylindocarbocyanine}
and switched to the confocal mode of excitation with linearly
polarized light. This allowed us to select a dipole oriented in the
substrate plane. In this configuration we obtained the image shown
in fig.~\ref{cr-scan-2}a) where complete quenching occurs within a
region of $160$ nm as the tip meets the molecule. The remaining
signal (see fig.~\ref{cr-scan-2}b)) is caused by residual
luminescence from the polymer and the tip. The interference pattern
is now absent in the confocal excitation.

The simple model of a Mie particle can be used to explain these
observations. We again choose a sphere of $100$ nm but now with the
dielectric constant of chromium. The results are displayed by the
solid curves in fig.~\ref{tip-theo-1}a) and b) for radially and in
fig.~\ref{tip-theo-2}a) and b) for tangentially oriented dipoles,
respectively. In the first case we note a moderate excitation
enhancement up to $K \approx 7$ at a separation of $10$ nm. At the
same time the quantum yield drops to $\eta \approx 20\%$ and thus we
estimate an overall fluorescence enhancement on the order of $S_f
\approx 1.4$. Already at a distance of $19$ nm the ratio between
$\gamma_{rad}$ and $\gamma_{nr}$ reverses in favor of the
non-radiative decay, diminishing $\eta$. Depending on the precise
distance, $S_f$ may rise above or drop below $S_0$. The situation is
different for the tangential orientation where both $K$ and $\eta$
fall below their initial values at any distance $z$. Particularly,
at $z \approx 10$~nm we estimate $S_f \approx
S_0/80$~\cite{Kuehn:06b}. The interference between the incident and
the scattered fields is now destructive on the surface of the sphere
so that the resulting tangential field nearly vanishes as would be
expected for a perfect metal. By the same token radiation from the
molecule is reduced close to the sphere. This inhibition does not
necessarily mean quenching but merely a longer lifetime although in
the presence of non-radiative decay there will always be quenching.

In the last two sections we have considered the effect of a sharp dielectric or metallic object on the fluorescence of a
single molecule. In each case, the orientation of the molecule with respect to the tip axis plays a central role. In the
tangential orientation fluorescence will always drop in the near field of a tip whereby its magnitude will depend on the
presence of absorption in the tip. For the vertical orientation, enhancement will occur for dielectric tips, but quenching
will extinguish the fluorescence at small separations from metallic tips. The tips discussed so far can be considered as
lightning rod antennas without any resonances.

\subsection{Single molecules and metallic nanoparticles as resonant dipole antennas}

Tips and even lithographically fabricated nanostructures vary substantially from one realization to the next. Furthermore,
it is a very demanding task to characterize a given nanostructure so that one can account for its geometrical features in
a theoretical model. Finally, extended tips are not expected to have well-defined plasmon resonances in the optical
domain. To get around these difficulties, a few years ago we initiated a series of experiments where individual colloidal
particles are carefully selected and examined via plasmon resonance spectroscopy and
tomography~\cite{Kalkbrenner:01,Kalkbrenner:04}. By attaching a single gold particle to the end of a glass fiber tip, we
could examine the modification of its plasmon spectrum close to surfaces~\cite{Kalkbrenner:05,Buchler:05} and investigate
the influence of the particle on the fluorescence of single fluroescent molecules~\cite{Kuehn:06}. In the next sections,
we discuss our results from the latter system. In particular, we provide direct studies of the enhancement of the
excitation at particle plasmon resonance, modification of the molecular fluorescence lifetime and its emission spectrum,
and the tip's strong influence on the emission pattern of the molecule.

\begin{figure}\centering
\includegraphics[width = 6 cm]{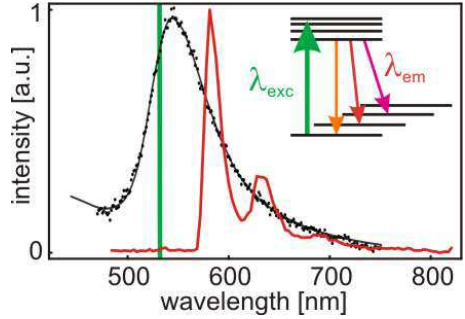}
\caption{The plasmon resonance of the nanoparticle (\textit{black})
partially overlaps the emission spectrum of a terrylene molecule
(\textit{red}). Possible emission pathways are illustrated in a
Jablonski-diagram. The green line indicates the excitation
wavelength.} \label{gnp-scatt-fluor}
\end{figure}

\subsubsection{Fluorescence enhancement and quenching}

Gold particles of $a=50$ nm were attached to glass tips following the procedure of Ref.~\cite{Kalkbrenner:01}. We
repeatedly monitored the scattering spectrum (see fig.~\ref{gnp-scatt-fluor}) during the experiment as a sensitive
indicator of the well-being of the tip. Single molecules were identified in a wide-field excitation arrangement and using
a sensitive CCD camera. An example of the radiation pattern of a single molecule is shown in fig.~\ref{setup}c).

\begin{figure} \centering
\includegraphics[width= 8 cm]{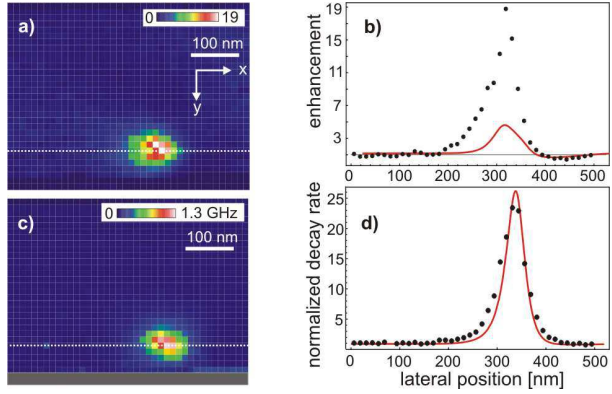}
\caption{Single molecule near-field image with a resonantly excited
gold-nanoparticle. a) Map of the fluorescence intensity $S_f$. b)
Cross-section from a). c) Corresponding map of the excited state
decay rate $\gamma$. d) Cross-section from c). Calculations of
$S_f/S_0$ and $\gamma/\gamma^0$ are also displayed as solid red
lines in (b) and (d).} \label{lat-scan}
\end{figure}

Once a molecule was selected, its fluorescence was directed to an avalanche photodiode and the gold particle was scanned
close to the sample surface. The near-field images of a single molecule shown in fig.~\ref{lat-scan} were obtained with
pulsed excitation light of $532$~nm. The fluorescence intensity $S_f$ typically exhibits a sharp rise over a region of $60
\times 60$~nm$^2$ as the tip comes close to the molecule. From the signal $S_0$ of the unperturbed molecule (blue shade)
we deduce an enhancement factor of 19 (fig.~\ref{lat-scan}b)) after carefully subtracting the residual luminescence $S_b$
from the tip and the sample substrate~\cite{Kuehn:06}. In other measurements (over many months, including many samples and
many tips) we have obtained enhancement factors ranging between 10 and 32. These variations are caused by different depths
$z_{min}$ of the molecules in the film (see below). We point out a slight asymmetry in the peak shape and the slight
suppression of the signal at $x \approx 440$ nm. These are due to the tilt of the molecule and to interference effects
discussed in Ref.~\cite{Kuehn:06} and its EPAPS Supplementary Material.

An important advantage of our experiment is that we have also
recorded the molecular fluorescence lifetime at every scan pixel
simultaneously as the fluorescence intensity. The corresponding map
of the decay rate is shown in fig.~\ref{lat-scan}c). Again a region
of accelerated decay can be seen with similar lateral extensions.
The decay rate $\gamma=\tau^{-1}$ is found to peak at about 22 times
the unperturbed value. We again note a slight asymmetry in the
cross-section shown in fig.~\ref{lat-scan}d) produced by the tilt of
the molecular dipole moment\cite{Frey2004,Kuehn:06}.

\begin{figure} \centering
\includegraphics[width = 8 cm]{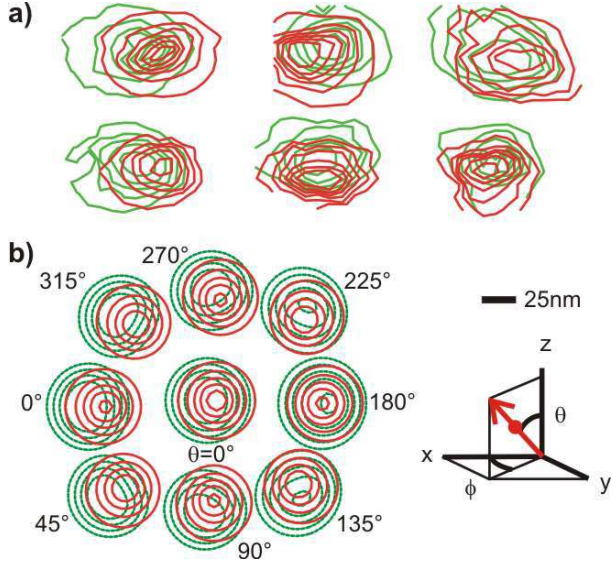}
\caption{a) Contours of constant enhancement (green) and decay rate
(red) for five different molecules. b) Calculated contours for
different azimuthal orientations of the molecular dipole $\phi$ at
$\theta = 15^\circ$ as defined in the inset.} \label{contour}
\end{figure}

A close examination of the scan images in figs.~\ref{lat-scan}a) and
c) reveals a slight systematic offset between the two enhancement
spots. To help the visualization of this interesting effect, in
fig.~\ref{contour}a) we show the superposed contour plots of these
two quantities from several molecules but the same gold
nanoparticle. The points of highest enhancement and fastest decay
rate are typically displaced from one another by up to $\Delta x
=20$ nm. The direction of the displacement, however, varies for
different molecules. We have presented the detailed analysis of this
issue in the supplementary materials of Ref.~\cite{Kuehn:06}. In a
nutshell, as the particle moves across the molecule, the relative
orientation of the molecular dipole changes with respect to the gold
sphere. Since tangential and radial orientations result in
completely different excitation enhancement and fluorescence
lifetime modification behaviors, a slight tilt of the molecule
(about $15^\circ$ for terrylene in pT~\cite{Pfab:04}) causes a
notable asymmetry between the intensity and lifetime images.

As discussed in the supplement of Ref.~\cite{Kuehn:06}, having a good estimate of the separation between the molecule and
the particle and knowing the tilt of the molecular dipole, we can calculate the modification of the excitation intensity
at the position of the molecule, of its radiative and nonradiative decay rates and thus its quantum efficiency. Based on
these calculations, we have plotted the expected profile of the detected fluorescence by the red curve in
fig.~\ref{lat-scan}b). The spatial extent of the enhancement region and its asymmetric shape agree well with the
predictions of theory apart from a slight broadening in the wings of the experimental curves. However, the magnitude of
$S_f$ falls short by a factor of 2-3 compared to the experimental values. This implies that $K$ and/or $\eta$ are larger
for the true geometry than in the simplified model~\cite{Kuehn2007}. We point out, however, that an accurate description
of the system must account for pT-air interface~\cite{Weeber1999, Anger2006} as well as the evanescent mode of excitation
for the tip.

\begin{figure} \centering
\includegraphics[width = 8 cm]{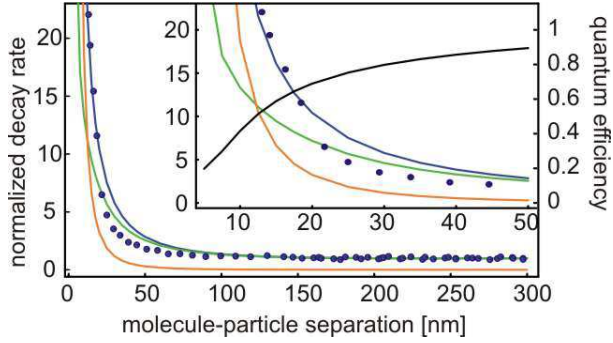}
\caption{Fluorescence decay rate for a $\bf \hat z$ approach. The symbols show the experimental decay rate $\gamma _{tot}$
and the lines give the calculated values of $\gamma_{rad}$ (\textit{orange}), $\gamma_{nr}$ (\textit{red}) and
$\gamma_{tot}$ (\textit{blue}). Inset: magnified plot with quantum yield $\eta$ (\textit{black}).} \label{appr-scan}
\end{figure}

It is also instructive to examine $K$ and $\gamma$ as a function of
the molecule-particle separation in the $z$ direction. As shown by
the pink curve in fig.~\ref{appr-scan}b), calculations predict an
increase in both $\gamma_{rad}$ and $\gamma_{nr}$ as $z$ is
decreased, but $\gamma_{nr}$ is expected to dominate at very small
distances. The blue symbols in the experimental data show the
measured total decay rate as the particle was approached to the
molecule~\cite{Kuehn:06}. The agreement with the blue curve, which
is the sum of the pink and green plots, is very good except for a
slightly weaker decay around $z \approx 50$ nm. Nevertheless, it is
apparent that $\gamma_{nr}$ starts to dominate at $z \leq 12$ nm and
leads to a decline of $\eta$ (inset). Equality of $\gamma_{rad}$ and
$\gamma_{nr}$ marks the point of $\eta = 50\%$. At distances $\leq
5$ nm the signal $S_f$ is expected to fall below its original value
$S_0$. In this experiment, the tip could not be approached closer to
the molecule because of the finite depth of the latter in the
sample. In fact, in our system the great majority of molecules that
are very close to the upper surface tend to bleach, most probably
due to residual or inter-diffusing oxygen~\cite{Christ2001,
Renn2006}.

\begin{figure} \centering
\includegraphics[width = 7.5 cm]{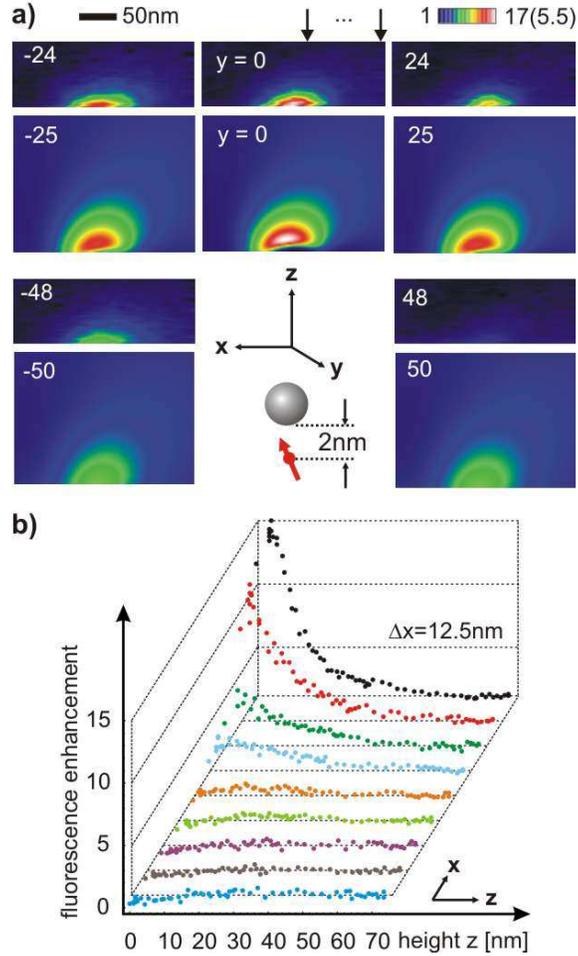}
\caption{a) First and third row: $xz$ section scans of $S_f$ at
different constant coordinates $y$ as given above (in nm). Second
and fourth row: calculation of $S_f(x, z, y=const)$ for a dipole
that is tilted by angle $\alpha = 25^\circ$ toward $\bf \hat x$ and
$z_{min} = 2 nm$. b) A selection of vertical cross sections from the
$y = 0$ image, as indicated by the arrows in a).} \label{appr-sec}
\end{figure}

In order to explore the complete three dimensional near-field
interaction of a gold particle and a single molecule, we have also
scanned the tip in the $xz$ plane at different $y$`s. The
experimental data is presented in the first and the third rows of
fig.~\ref{appr-sec}a). Here $S_f$ reaches an absolute maximum
enhancement of 17 when the tip assumes the smallest separation
$z_{min}$ from the sample surface for $y=0$. A selection of approach
curves from the $y=0$ data is shown in fig.~\ref{appr-sec}b). When
the $x$-separation between the particle and the molecule is large
(lower graphs), only a slight quenching occurs for small $z$. As the
particle comes closer to the molecule in the $x$ direction (top
graphs), the fluorescence is first enhanced but it then reaches a
maximum and turns around at about $5$ nm. This effect has been also
reported in Ref.~\cite{Anger:06}. We also note that for some $y$
values (e.g. $y=24$~nm), $S_f$ assumes its maximum value at a
distance $z> z_{min}$  (see fig.~\ref{appr-sec}a)). This phenomenon
is more clearly seen in the calculations that are plotted in the
second and fourth rows of fig.~\ref{appr-sec}a) below each
corresponding experimental image. Furthermore, the skewed elongated
character of the experimental regions of enhancement is reproduced
in the calculations with a nearly quantitative agreement. We
mention, however, that in these calculations a better agreement was
obtained with the experimental data when we chose an inclination
angle of $\alpha \approx 25^\circ$ toward $\bf \hat x$.

\begin{figure} \centering
\includegraphics[width= 8 cm]{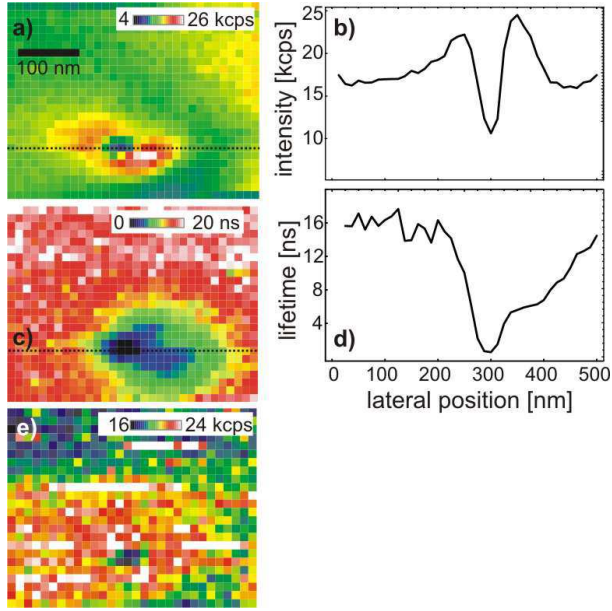}
\caption{a) Fluorescence signal $S_f$ for a near-field scan with a
gold nano-particle at minimal $z_{min}$ and excitation wavelength
$\lambda_{exc} = 532$ nm. b) Cross-section from a). c) Map of the
lifetime $\tau$ recorded simultaneously with a). d) Cross-section
from c). e) Same as a) but with $\lambda_{exc} = 488$ nm.}
\label{quench-green}
\end{figure}

The data in fig.~\ref{appr-sec} let us predict that for a normally
oriented molecule and a small tip-sample separation, we can expect a
dip in the center of the image due to a pronounced quenching (see
the $y=0$ data set in the second row of fig.~\ref{appr-sec}a)). For
large $z_{min}$ or tilted molecules the enhancement will have a
central maximum as in fig.~\ref{lat-scan} and possibly a weak
minimum. The data in fig.~\ref{quench-green} show lateral scans of
$S_f$ and $\tau$ at very small particle-molecule separations where
quenching becomes very important. A significant drop of about $50\%$
in the fluorescence rate $S_f$ (see fig.~\ref{quench-green}b))
coincides with a drop in the lifetime $\tau$ from $16$~ns to below
$1$ ns in fig.~\ref{quench-green}d). The central quenching pit in
fig.~\ref{quench-green}a) is flanked by a circular region where the
fluorescence is enhanced by 1.5 times. The corresponding map of the
lifetime shown in fig.~\ref{quench-green}c) and d) exhibits a steep
rise to the left and a slower recovery with a shoulder to the right
from the main quenching spot. The fact that both lifetime and
fluorescence signals drop is indicative of quenching. We mentions
that a small dip of the fluorescence enhancement at the center of
lateral scans has been also reported in Ref.~\cite{Anger:06} but
corresponding lifetime measurements were missing.

The strong enhancement discussed here was explained via the optical
properties of a gold nanoparticle using the same formalism as for a
glass or chromium spheres treated earlier. An important and decisive
feature of a gold nanoparticle is that it supports a plasmon
resonance, making it behave like a ''resonant optical
antenna''~\cite{Alda2005, Crozier2003, Greffet2005}. In the next two
sections we discuss the role of the plasmon resonance on the
molecular excitation and emission processes.

\subsubsection{The role of plasmon resonance in the molecular excitation}

\begin{figure} \centering
\includegraphics[width = 8 cm]{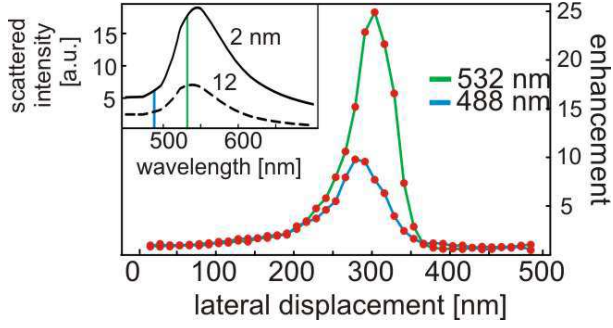}
\caption{Comparison of the enhancement at resonant (\textit{green})
and non-resonant (\textit{blue}) excitations. The inset shows a
calculation of $K(\lambda_{exc})$ at distances of $2$ and $12$ nm.}
\label{green-blue}
\end{figure}

As shown by the spectrum in fig.~\ref{gnp-scatt-fluor}, the plasmon
resonances of the nanoparticles we studied show a maximum around
$550$ nm but a minimal scattering response close to $490$~nm. This
means that also $K$ should depend on the wavelength of excitation.
We have compared the fluorescence signals $S_f$ for the excitation
with blue ($488$ nm) and green ($532$ nm) light. To keep all other
parameters constant, we combined the two colors on the same optical
path and switched between the sources during a raster scan. The
images thus acquired resemble those of fig.~\ref{lat-scan}. However,
the enhancement with the blue light is smaller by a factor of 2.5 as
can be seen from a line scan shown in
fig.~\ref{green-blue}~\cite{Kuehn:06}. The difference between the
maxima of $S_f$ is fully attributed to $K$ since the conditions for
the emission of the molecule remain unaltered. The inset shows
calculations for $K(\lambda)$ or equivalently, the near-field
plasmon resonance at $z=2$ and $12$ nm. An enhancement ratio of 2.5
corresponds to $z=8$~nm.

Another interesting effect is shown in fig.~\ref{quench-green}e) where the fluorescence intensity of the same molecule as
in fig.~\ref{quench-green}a) is plotted but this time under excitation at $488$~nm. Under the weaker excitation at this
wavelength the circular region of enhancement that was observed for green excitation is absent and only the central
quenching pit remains.

\subsubsection{The role of plasmon resonance in the molecular emission Spectrum}

According to Fermi Golden Rule $\gamma_{rad}(\omega)$ is
proportional to the photonic density of states
$\rho(\omega)$~\cite{Joulain2003, Klimov2001}. Similar to the case
of an emitter placed close to a perfect mirror, the density of
states and thus $\gamma_{rad}$ are expected to be modified for
emitters close to resonant
nanostructures~\cite{Metiu1984,Kaell2005,Rogobete:07}. Above, we
have provided experimental data for the change of the total
fluorescence decay rate and have discussed theoretical calculations
that help us estimate how much of this observed change is due to
$\gamma_{rad}$ as opposed to $\gamma_{nr}$. A direct experimental
measurement of these two parameters remains the topic of a future
publication~\cite{Kuehn2007}. However, here, we give direct evidence
for the modification of $\gamma_{rad}$ through the change of the
emission spectrum.

\begin{figure} \centering
\includegraphics[width = 8 cm]{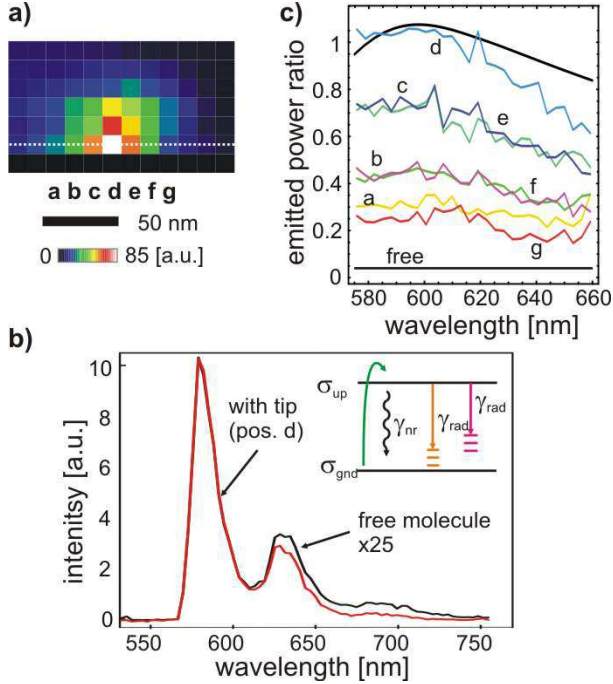}
\caption{a) Map of the integrated fluorescence intensity as the gold
particle is scanned across a terrylene molecule. b) Normalized
fluorescence spectra of the unperturbed molecule (black) and at
maximal enhancement (pixel ''d'' of the line indicated in a). c)
Ratio of the spectra measured at pixels a-g to the unperturbed
molecule.} \label{emi-spec}
\end{figure}

To investigate the modification of the molecular emission spectrum
under the influence of a gold nanoparticle, we have recorded the
fluorescence spectrum of a single molecule using a grating
spectrometer as the tip was scanned across the molecule. The
integrated spectra are shown in fig.~\ref{emi-spec}a), constituting
a total near-field image $S_f$. In these measurements, the
enhancement reached a factor of 25 just before the molecule
underwent photobleaching. In fig.~\ref{emi-spec}b) we show the
normalized spectrum for the most intense pixel labeled ''d'' in
fig.~\ref{emi-spec}a) (red curve) and the spectrum of the same
unperturbed molecule (black curve). It is evident that the enhanced
spectrum falls below the reference spectrum at longer wavelengths
($630 - 750$ nm).

The fluorescence spectrum of terrylene in fig.~\ref{emi-spec} shows
three peaks due to the emission from the lowest vibrational state of
the electronic upper state to different vibrational levels of the
electronic ground state (see the inset in
fig.~\ref{gnp-scatt-fluor}). The spectral power density $p(\lambda)$
is proportional to $\sigma_{up} \gamma_{rad}(\lambda)$ where
$\sigma_{up}$ denotes the common upper state population. Although we
do not know $\sigma_{up}$ in the absence and presence of the gold
particle, because it is the same for all emission channels, we can
probe the effect of $\gamma_{rad}(\lambda)$ by comparing the
fluorescence enhancement at different wavelengths. To do this, we
have computed the ratio
$\gamma_{rad}^{gold}(\lambda)/\gamma_{rad}^{free}(\lambda)$ by
dividing the spectra obtained with and without the gold particle.
This ratio is shown in fig.~\ref{emi-spec}c) for a selection of
pixels in fig.~\ref{emi-spec}a). The imbalance in the emission
enhancement increases gradually as the distance between the molecule
and nanoparticle decreases. At the maximum of fluorescence intensity
(curve d in fig.~\ref{emi-spec}c) we find that the highest emission
enhancement is assumed around $600$ nm, i.e. red-shifted from the
scattering resonance at $550$ nm~\cite{Anger2006, Rogobete:07}. The
overall tendency of the experimental data is in a reasonable
agreement with the theoretical ratio of
$\gamma_{rad}^{gold}(\lambda)/\gamma_{rad}^{free}(\lambda)$ plotted
by the black curve in fig.~\ref{emi-spec}c).

From these results we can conclude that it is crucial to consider the optical properties of the material in the design of
an optical antenna~\cite{Crozier2003, Calander2002}. The material non-idealities compared to radio frequencies alter the
relation between the antenna resonance and shape. Excitation enhancement and emission efficiency cannot be related by
simple reciprocity arguments but have to account for absorption~\cite{Thomas2004}.

\subsubsection{Radiation pattern}

Antenna design aims at connecting \textit{localized} sources and
detectors through specific directional characteristics. A dipole,
however, radiates nearly isotropically. Efficient excitation and
collection requires high-grade optics that are not viable in many
applications. To improve the directionality from a molecule, it has
been proposed that reflectors and directors using the Yagi-Uda
principle can be used~\cite{Li2007, Hofmann2007}. It turns out that
a single nanoparticle can also alter the radiation pattern of the
molecule. Indeed, the modification of angular emission due to a
solid near-field tip has been reported~\cite{Gersen2000}.

To investigate the radiation pattern from a single molecule in the
presence of a nano-antenna, we directed the emitted light at high
magnification onto an imaging CCD camera. The beam produced by an
abberation-free microscope objective is the transform of the
radiation pattern $I(\theta, \phi)$ at the focal plane according to
$I(\rho, \phi) \propto I(\theta, \phi)/\cos{\theta}$ with $\rho = f
\sin{\theta}$ where $f$ is the effective focal
length~\cite{Lieb2004}. An axially oriented dipole thus produces a
radially polarized doughnut-shaped distribution~\cite{Bohmer2004,
Pfab:04} (see fig.~\ref{setup}d). In case of terrylene embedded in
para-terphenyl there exists also a slight asymmetry due to its
tilt~\cite{Pfab:04}.

\begin{figure} \centering
\includegraphics[width = 6 cm]{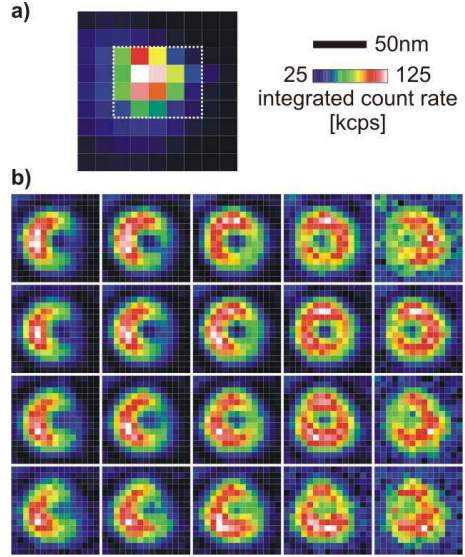}
\caption{a) Integrated intensity map for a near-field scan of a gold
particle across a terrylene molecule. b) Radiation pattern on the
CCD camera for a set of displacements corresponding to the 20 pixels
within the white box in a).} \label{ccd-pat}
\end{figure}

We have collected a sequence of such images as the nanoparticle was
scanned across a molecule. As before, the integrated count rate on
the camera produces a typical enhancement map shown in
fig.~\ref{ccd-pat}a), with a maximum of 25 in this scan.
Figure~\ref{ccd-pat}b) shows the spatially resolved camera images
for the tip locations corresponding to the pixels within the white
square in a). Inspection of these images reveals a redistribution of
the intensity going from a ``waning'' to a ``waxing'' crescent in
the first and last columns, respectively. Similarly, for the top and
bottom rows the bright regions can be found in the upper and lower
sections of the doughnut. A simple trend is obvious: the weight of
the intensity in the pattern follows the direction of displacement
of the tip.

\begin{figure} \centering
\includegraphics[width=8 cm]{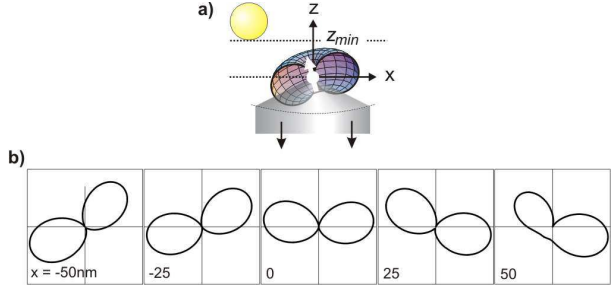}
\caption{Calculated radiation pattern for the dipole - gold sphere
system. a) Model schematic with the parameters: $\alpha = 15^\circ$,
$z_{min} = 10$ nm. b) Radiation pattern of a molecule in the
presence of a gold sphere with displacement $x = (-50...50)$ nm. The
total power emitted at any point is given by the intensity map
according to fig.~\ref{lat-scan}.}\label{pat-sim}
\end{figure}

To explain the redistribution of power, we have calculated the
radiation pattern for a dipole close to a gold
sphere~\cite{Chew1987} at different locations. Cross-sections of
this pattern in the $xz$-plane are shown in fig.~\ref{pat-sim}. The
model predicts an inclination of the radiation pattern towards the
line connecting sphere and molecule~\cite{Thomas2004}. This is a
consequence of the radial component of the molecular dipole which
polarizes the sphere to produce a large common dipole moment. The
emission from the tangential component remains inferior in our
geometry. It merely produces a slight right-left asymmetry and most
notably the alleviation of the pinch-off at $x=50$ nm. We note that
our calculation does not represent the true geometry including the
pT interface~\cite{Hellen1987, Arnoldus2004}, which directs almost
the entire radiation into the higher index medium. The change of the
emission pattern shows the necessity for a high collection
solid-angle and implies that the excitation conditions should be in
principle optimized for each given separation $\bf r$. By using of
the reciprocity theorem, we can predict that the excitation beam
should be shaped similar to the radiation pattern. Optimal
excitation can therefore be achieved with a radially polarized
doughnut beam \cite{Quabis2000, Anger2006}.

\subsection{Optimizing antenna structures}

In the previous sections, we have shown that a gold nanoparticle
acts as a nano-antenna to enhance the fluorescence signal of single
emitters. However, we also showed that absorption of light by the
gold particle causes quenching so that one cannot harvest the
greatest excitation and spontaneous emission enhancements at very
small separations. The question that arises is then if one could
devise an arrangement for achieving a very high spontaneous emission
rate without sacrificing the quantum efficiency. Furthermore, it
would be desirable to achieve more significant enhancement of
excitation than was possible from a nanosphere. We have recently
tackled this problem theoretically for simple antenna geometries of
two nanoparticles~\cite{Rogobete:07}. Using full electrodynamics
simulations~\cite{Kaminski2007}, we have applied some simple design
rules to nanoantennae made of two prolate spheroidal gold particles
as sketched in fig.~\ref{mario}a). Due to the lack of space, we
refer the interested reader to Ref.~\cite{Rogobete:07} for a
detailed discussion of this topic. Here it suffices to state that by
varying size, aspect ratio, distance and background medium, we have
found that quenching can be avoided to a large extent. Moreover,
radiative decay enhancements up to three order of magnitudes can be
achieved in the near-infrared spectral range~\cite{Rogobete:07,
Agio2007, Rogobete2007a}. By invoking reciprocity, we conclude that
this enhancement also applies to the near field and thus to the
excitation process, as previously pointed out in the
literature~\cite{Aravind1981}.

\begin{figure}\centering
\includegraphics[width = 8 cm]{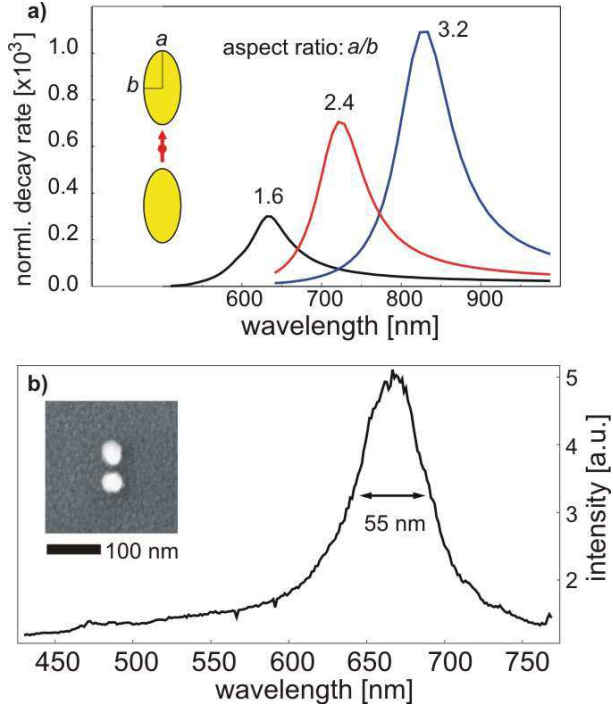} \caption{\label{mario} a) Normalized radiative decay rate as a
function of the NPs aspect ratio $a/b$ for an emitter coupled to the
gold nanoantenna as shown in the inset. NPs details: aspect ratio
1.6 ($a$=76 nm, $b$=48 nm), 2.4 ($a$=100 nm, $b$=42 nm), and 3.2
($a$=122 nm, $b$=32 nm). The NPs are separated by about 20 nm and
are embedded in a medium with refractive index equal to 1.3. b) The
measured plasmon spectrum of a gold nanoantenna fabricated by
electron beam lithography. The incident light is polarized parallel
to the nanoantenna long axis. Inset: SEM image of the fabricated
sample.}\label{mario}
\end{figure}

Although we expect the best results for the near-infrared emitters,
it is possible to optimize the coupling of gold nanoparticles to
emitters in the visible range. Figure~\ref{mario}a) describes how
the aspect ratio of the nanoparticles can be used to adjust the
plasmon resonance to the desired wavelength. The quantum efficiency
is close to one and insensitive to the aspect ratio for wavelengths
larger than about 750 nm, while it exhibits a rapid drop for shorter
wavelengths, depending on the nanoparticle's shape and background
medium~\cite{Agio2007, Rogobete2007a}. Similar trends should be
found also for gold nanorods~\cite{Aizpurua2005}.

To close, we briefly present an experimental progress report on the
fabrication of double-particle antennae~\cite{Agio2007}. The inset
in fig.~\ref{mario}b) displays a scanning electron microscope image
of a nanoantenna fabricated by electron-beam lithography. Such a
structure with an overall size around 100~nm is very promising for
the realization of a hybrid subwavelength emitter with tailored
radiative features. The main part of fig.~\ref{mario}b) shows the
plasmon resonance recorded from such a structure~\cite{Agio2007}.
The resonance at 670 nm corresponds to slightly elongated
nanoparticles separated by only about 15 nm. From the experimental
point of view, the fabrication of the ideal nanoantennae remains a
nontrivial task. Furthermore, proper positioning of a single
molecule in the gap between two nanoparticles is quite challenging.
Nevertheless, there is no doubt that in the next few years
nano-optics will witness a rapid progress in this direction.

\section*{Acknowledgements}
We thank F. Robin for fabricating gold nanoantennae at FIRST, the Center for Micro- and Nanoscale Science at ETH Zurich.
We thank U. H\aa kanson, L. Rogobete and F. Kaminski for fruitful discussions. This work was supported by the Swiss
Ministry of Education and Science (EU IP-Molecular Imaging), the ETH Zurich initiative on Composite Doped Metamaterials
(CDM) and the Swiss National Foundation (SNF).


\begin{thebibliography}{105}
\expandafter\ifx\csname
natexlab\endcsname\relax\def\natexlab#1{#1}\fi
\expandafter\ifx\csname bibnamefont\endcsname\relax
  \def\bibnamefont#1{#1}\fi
\expandafter\ifx\csname bibfnamefont\endcsname\relax
  \def\bibfnamefont#1{#1}\fi
\expandafter\ifx\csname citenamefont\endcsname\relax
  \def\citenamefont#1{#1}\fi
\expandafter\ifx\csname url\endcsname\relax
  \def\url#1{\texttt{#1}}\fi
\expandafter\ifx\csname urlprefix\endcsname\relax\def\urlprefix{URL
}\fi \providecommand{\bibinfo}[2]{#2}
\providecommand{\eprint}[2][]{\url{#2}}

\bibitem[{\citenamefont{Sommerfeld}(1909)}]{Sommerfeld1909}
\bibinfo{author}{\bibfnamefont{A.}~\bibnamefont{Sommerfeld}},
  \bibinfo{journal}{Ann. Phys. (Leipzig)} \textbf{\bibinfo{volume}{28}},
  \bibinfo{pages}{665–736} (\bibinfo{year}{1909}).

\bibitem[{\citenamefont{Purcell}(1946)}]{Purcell:46}
\bibinfo{author}{\bibfnamefont{E.~M.} \bibnamefont{Purcell}},
  \bibinfo{journal}{Phys. Rev.} \textbf{\bibinfo{volume}{69}},
  \bibinfo{pages}{681} (\bibinfo{year}{1946}).

\bibitem[{\citenamefont{Drexhage}(1974)}]{Drexhage:74a}
\bibinfo{author}{\bibfnamefont{K.~H.} \bibnamefont{Drexhage}},
  \bibinfo{journal}{Progress in Optics} \textbf{\bibinfo{volume}{12}},
  \bibinfo{pages}{165} (\bibinfo{year}{1974}).

\bibitem[{\citenamefont{Berman}(1994)}]{Berman}
\bibinfo{author}{\bibfnamefont{P.~R.} \bibnamefont{Berman}},
  \emph{\bibinfo{title}{Cavity Quantum Electrodynamics}}
  (\bibinfo{publisher}{Academic Press}, \bibinfo{year}{1994}).

\bibitem[{\citenamefont{Chang et~al.}(1996)\citenamefont{Chang, Fann, and
  Lin}}]{Chang03}
\bibinfo{author}{\bibfnamefont{R.}~\bibnamefont{Chang}},
  \bibinfo{author}{\bibfnamefont{W.}~\bibnamefont{Fann}}, \bibnamefont{and}
  \bibinfo{author}{\bibfnamefont{S.~H.} \bibnamefont{Lin}},
  \bibinfo{journal}{Appl. Phys. Lett.} \textbf{\bibinfo{volume}{69}},
  \bibinfo{pages}{2338} (\bibinfo{year}{1996}).

\bibitem[{\citenamefont{Henkel and Sandoghdar}(1998)}]{henkel:98}
\bibinfo{author}{\bibfnamefont{C.}~\bibnamefont{Henkel}} \bibnamefont{and}
  \bibinfo{author}{\bibfnamefont{V.}~\bibnamefont{Sandoghdar}},
  \bibinfo{journal}{Opt. Comm.} \textbf{\bibinfo{volume}{158}},
  \bibinfo{pages}{250} (\bibinfo{year}{1998}).

\bibitem[{\citenamefont{Rogobete et~al.}(2003)\citenamefont{Rogobete, Schniepp,
  Sandoghdar, and Henkel}}]{Rogobete:03}
\bibinfo{author}{\bibfnamefont{L.}~\bibnamefont{Rogobete}},
  \bibinfo{author}{\bibfnamefont{H.}~\bibnamefont{Schniepp}},
  \bibinfo{author}{\bibfnamefont{V.}~\bibnamefont{Sandoghdar}},
  \bibnamefont{and} \bibinfo{author}{\bibfnamefont{C.}~\bibnamefont{Henkel}},
  \bibinfo{journal}{Opt. Lett.} \textbf{\bibinfo{volume}{28}},
  \bibinfo{pages}{1736} (\bibinfo{year}{2003}).

\bibitem[{\citenamefont{Koenderink et~al.}(2005)\citenamefont{Koenderink,
  Kafesaki, Soukoulis, and Sandoghdar}}]{Koenderink:05b}
\bibinfo{author}{\bibfnamefont{A.~F.} \bibnamefont{Koenderink}},
  \bibinfo{author}{\bibfnamefont{M.}~\bibnamefont{Kafesaki}},
  \bibinfo{author}{\bibfnamefont{C.~M.} \bibnamefont{Soukoulis}},
  \bibnamefont{and}
  \bibinfo{author}{\bibfnamefont{V.}~\bibnamefont{Sandoghdar}},
  \bibinfo{journal}{Opt. Lett.} \textbf{\bibinfo{volume}{30}},
  \bibinfo{pages}{3210} (\bibinfo{year}{2005}).

\bibitem[{\citenamefont{Rogobete
  et~al.}(2007{\natexlab{a}})\citenamefont{Rogobete, Kaminski, Agio, and
  Sandoghdar}}]{Rogobete:07}
\bibinfo{author}{\bibfnamefont{L.}~\bibnamefont{Rogobete}},
  \bibinfo{author}{\bibfnamefont{F.}~\bibnamefont{Kaminski}},
  \bibinfo{author}{\bibfnamefont{M.}~\bibnamefont{Agio}}, \bibnamefont{and}
  \bibinfo{author}{\bibfnamefont{V.}~\bibnamefont{Sandoghdar}},
  \bibinfo{journal}{Opt. Lett.} \textbf{\bibinfo{volume}{32}},
  \bibinfo{pages}{1623} (\bibinfo{year}{2007}{\natexlab{a}}).

\bibitem[{\citenamefont{Schniepp and Sandoghdar}(2002)}]{Schniepp:02}
\bibinfo{author}{\bibfnamefont{H.}~\bibnamefont{Schniepp}} \bibnamefont{and}
  \bibinfo{author}{\bibfnamefont{V.}~\bibnamefont{Sandoghdar}},
  \bibinfo{journal}{Phys. Rev. Lett.} \textbf{\bibinfo{volume}{89}},
  \bibinfo{pages}{257403} (\bibinfo{year}{2002}).

\bibitem[{\citenamefont{K\"{u}hn et~al.}(2006)\citenamefont{K\"{u}hn,
  H{\aa}kanson, Rogobete, and Sandoghdar}}]{Kuehn:06}
\bibinfo{author}{\bibfnamefont{S.}~\bibnamefont{K\"{u}hn}},
  \bibinfo{author}{\bibfnamefont{U.}~\bibnamefont{H{\aa}kanson}},
  \bibinfo{author}{\bibfnamefont{L.}~\bibnamefont{Rogobete}}, \bibnamefont{and}
  \bibinfo{author}{\bibfnamefont{V.}~\bibnamefont{Sandoghdar}},
  \bibinfo{journal}{Phys. Rev. Lett.} \textbf{\bibinfo{volume}{97}},
  \bibinfo{pages}{017402} (\bibinfo{year}{2006}).

\bibitem[{\citenamefont{K\"{u}hn and Sandoghdar}(2006)}]{Kuehn:06b}
\bibinfo{author}{\bibfnamefont{S.}~\bibnamefont{K\"{u}hn}} \bibnamefont{and}
  \bibinfo{author}{\bibfnamefont{V.}~\bibnamefont{Sandoghdar}},
  \bibinfo{journal}{Appl. Phys. B} \textbf{\bibinfo{volume}{84}},
  \bibinfo{pages}{211} (\bibinfo{year}{2006}).

\bibitem[{\citenamefont{Seelig et~al.}(2007)\citenamefont{Seelig, Leslie, Renn,
  Kuhn, Jacobsen, vandeCorput, Wyman, and Sandoghdar}}]{Seelig2007}
\bibinfo{author}{\bibfnamefont{J.}~\bibnamefont{Seelig}},
  \bibinfo{author}{\bibfnamefont{K.}~\bibnamefont{Leslie}},
  \bibinfo{author}{\bibfnamefont{A.}~\bibnamefont{Renn}},
  \bibinfo{author}{\bibfnamefont{S.}~\bibnamefont{Kuhn}},
  \bibinfo{author}{\bibfnamefont{V.}~\bibnamefont{Jacobsen}},
  \bibinfo{author}{\bibfnamefont{M.}~\bibnamefont{van de Corput}},
  \bibinfo{author}{\bibfnamefont{C.}~\bibnamefont{Wyman}}, \bibnamefont{and}
  \bibinfo{author}{\bibfnamefont{V.}~\bibnamefont{Sandoghdar}},
  \bibinfo{journal}{Nano Lett.} \textbf{\bibinfo{volume}{7}},
  \bibinfo{pages}{685} (\bibinfo{year}{2007}).

\bibitem[{\citenamefont{Fleischmann et~al.}(1974)\citenamefont{Fleischmann,
  Hendra, and McQillan}}]{Fleischmann1974}
\bibinfo{author}{\bibfnamefont{M.}~\bibnamefont{Fleischmann}},
  \bibinfo{author}{\bibfnamefont{P.}~\bibnamefont{Hendra}}, \bibnamefont{and}
  \bibinfo{author}{\bibfnamefont{A.}~\bibnamefont{McQillan}},
  \bibinfo{journal}{Chem. Phys. Lett.} \textbf{\bibinfo{volume}{26}},
  \bibinfo{pages}{163} (\bibinfo{year}{1974}).

\bibitem[{\citenamefont{Jackson}(1999)}]{Jackson-book}
\bibinfo{author}{\bibfnamefont{D.}~\bibnamefont{Jackson}},
  \emph{\bibinfo{title}{Classical Electrodynamics}} (\bibinfo{publisher}{Wiley
  and Sons}, \bibinfo{year}{1999}).

\bibitem[{\citenamefont{Metiu}(1984{\natexlab{a}})}]{Metiu:84}
\bibinfo{author}{\bibfnamefont{H.}~\bibnamefont{Metiu}},
  \bibinfo{journal}{Prog. Surf. Sci.} \textbf{\bibinfo{volume}{17}},
  \bibinfo{pages}{153} (\bibinfo{year}{1984}{\natexlab{a}}).

\bibitem[{\citenamefont{Moskovits}(1985)}]{Moskovits:85}
\bibinfo{author}{\bibfnamefont{M.}~\bibnamefont{Moskovits}},
  \bibinfo{journal}{Rev. Mod. Phys.} \textbf{\bibinfo{volume}{57}},
  \bibinfo{pages}{783} (\bibinfo{year}{1985}).

\bibitem[{\citenamefont{Gersten and Nitzan}(1981)}]{Gersten1981}
\bibinfo{author}{\bibfnamefont{J.}~\bibnamefont{Gersten}} \bibnamefont{and}
  \bibinfo{author}{\bibfnamefont{A.}~\bibnamefont{Nitzan}},
  \bibinfo{journal}{J. Chem. Phys.}
  \textbf{\bibinfo{volume}{75}}, \bibinfo{pages}{1139} (\bibinfo{year}{1981}).

\bibitem[{\citenamefont{Ruppin}(1982)}]{Ruppin1982}
\bibinfo{author}{\bibfnamefont{R.}~\bibnamefont{Ruppin}}, \bibinfo{journal}{J.
  Chem. Phys.} \textbf{\bibinfo{volume}{76}}, \bibinfo{pages}{1681}
  (\bibinfo{year}{1982}).

\bibitem[{\citenamefont{Chew}(1988)}]{chew:88}
\bibinfo{author}{\bibfnamefont{H.}~\bibnamefont{Chew}}, \bibinfo{journal}{Phys.
  Rev. A} \textbf{\bibinfo{volume}{38}}, \bibinfo{pages}{3410}
  (\bibinfo{year}{1988}).

\bibitem[{\citenamefont{Kulzer and Orrit}(2004)}]{Kulzer:04}
\bibinfo{author}{\bibfnamefont{F.}~\bibnamefont{Kulzer}} \bibnamefont{and}
  \bibinfo{author}{\bibfnamefont{M.}~\bibnamefont{Orrit}},
  \bibinfo{journal}{Ann. Rev. Phys. Chem.} \textbf{\bibinfo{volume}{55}},
  \bibinfo{pages}{585} (\bibinfo{year}{2004}).

\bibitem[{\citenamefont{Klar et~al.}(1998)\citenamefont{Klar, Perner, Grosse,
  {v}on Plessen, Spirkl, and Feldmann}}]{klar:98}
\bibinfo{author}{\bibfnamefont{T.}~\bibnamefont{Klar}},
  \bibinfo{author}{\bibfnamefont{M.}~\bibnamefont{Perner}},
  \bibinfo{author}{\bibfnamefont{S.}~\bibnamefont{Grosse}},
  \bibinfo{author}{\bibfnamefont{G.}~\bibnamefont{{v}on Plessen}},
  \bibinfo{author}{\bibfnamefont{W.}~\bibnamefont{Spirkl}}, \bibnamefont{and}
  \bibinfo{author}{\bibfnamefont{J.}~\bibnamefont{Feldmann}},
  \bibinfo{journal}{Phys. Rev. Let.} \textbf{\bibinfo{volume}{80}},
  \bibinfo{pages}{4249} (\bibinfo{year}{1998}).

\bibitem[{\citenamefont{Kalkbrenner et~al.}(2004)\citenamefont{Kalkbrenner,
  H{\aa}kanson, and Sandoghdar}}]{Kalkbrenner:04}
\bibinfo{author}{\bibfnamefont{T.}~\bibnamefont{Kalkbrenner}},
  \bibinfo{author}{\bibfnamefont{U.}~\bibnamefont{H{\aa}kanson}},
  \bibnamefont{and}
  \bibinfo{author}{\bibfnamefont{V.}~\bibnamefont{Sandoghdar}},
  \bibinfo{journal}{Nano Lett.} \textbf{\bibinfo{volume}{4}},
  \bibinfo{pages}{2309} (\bibinfo{year}{2004}).

\bibitem[{\citenamefont{Lindfors et~al.}(2004)\citenamefont{Lindfors,
  Kalkbrenner, Stoller, and Sandoghdar}}]{Lindfors:04}
\bibinfo{author}{\bibfnamefont{K.}~\bibnamefont{Lindfors}},
  \bibinfo{author}{\bibfnamefont{T.}~\bibnamefont{Kalkbrenner}},
  \bibinfo{author}{\bibfnamefont{P.}~\bibnamefont{Stoller}}, \bibnamefont{and}
  \bibinfo{author}{\bibfnamefont{V.}~\bibnamefont{Sandoghdar}},
  \bibinfo{journal}{Phys. Rev. Lett.} \textbf{\bibinfo{volume}{93}},
  \bibinfo{pages}{037401} (\bibinfo{year}{2004}).

\bibitem[{\citenamefont{van Dijk et~al.}(2006)\citenamefont{van Dijk,
  Tchebotareva, Orrit, Lippitz, Berciaud, Lasne, Cognet, and
  Lounis}}]{vanDijk:06}
\bibinfo{author}{\bibfnamefont{M.~A.} \bibnamefont{van Dijk}},
  \bibinfo{author}{\bibfnamefont{A.~L.} \bibnamefont{Tchebotareva}},
  \bibinfo{author}{\bibfnamefont{M.}~\bibnamefont{Orrit}},
  \bibinfo{author}{\bibfnamefont{M.}~\bibnamefont{Lippitz}},
  \bibinfo{author}{\bibfnamefont{S.}~\bibnamefont{Berciaud}},
  \bibinfo{author}{\bibfnamefont{D.}~\bibnamefont{Lasne}},
  \bibinfo{author}{\bibfnamefont{L.}~\bibnamefont{Cognet}}, \bibnamefont{and}
  \bibinfo{author}{\bibfnamefont{B.}~\bibnamefont{Lounis}},
  \bibinfo{journal}{Phys. Chem. Chem. Phys.} \textbf{\bibinfo{volume}{8}},
  \bibinfo{pages}{3486 } (\bibinfo{year}{2006}).

\bibitem[{\citenamefont{Nie and Emory}(1997)}]{Nie:97}
\bibinfo{author}{\bibfnamefont{S.}~\bibnamefont{Nie}} \bibnamefont{and}
  \bibinfo{author}{\bibfnamefont{S.~R.} \bibnamefont{Emory}},
  \bibinfo{journal}{Science} \textbf{\bibinfo{volume}{275}},
  \bibinfo{pages}{1102} (\bibinfo{year}{1997}).

\bibitem[{\citenamefont{Kneipp et~al.}(1997)\citenamefont{Kneipp, Wang, Kneipp,
  Perelman, Itzkan, Dasari, and Feld}}]{kneipp:97}
\bibinfo{author}{\bibfnamefont{K.}~\bibnamefont{Kneipp}},
  \bibinfo{author}{\bibfnamefont{Y.}~\bibnamefont{Wang}},
  \bibinfo{author}{\bibfnamefont{H.}~\bibnamefont{Kneipp}},
  \bibinfo{author}{\bibfnamefont{L.~T.} \bibnamefont{Perelman}},
  \bibinfo{author}{\bibfnamefont{I.}~\bibnamefont{Itzkan}},
  \bibinfo{author}{\bibfnamefont{R.~R.} \bibnamefont{Dasari}},
  \bibnamefont{and} \bibinfo{author}{\bibfnamefont{M.~S.} \bibnamefont{Feld}},
  \bibinfo{journal}{Phys. Rev. Lett.} \textbf{\bibinfo{volume}{78}},
  \bibinfo{pages}{1667} (\bibinfo{year}{1997}).

\bibitem[{\citenamefont{Ambrose et~al.}(1994)\citenamefont{Ambrose, Goodwin,
  Martin, and Keller}}]{Ambrose:94}
\bibinfo{author}{\bibfnamefont{W.~P.} \bibnamefont{Ambrose}},
  \bibinfo{author}{\bibfnamefont{P.~M.} \bibnamefont{Goodwin}},
  \bibinfo{author}{\bibfnamefont{J.~C.} \bibnamefont{Martin}},
  \bibnamefont{and} \bibinfo{author}{\bibfnamefont{R.~A.}
  \bibnamefont{Keller}}, \bibinfo{journal}{Science}
  \textbf{\bibinfo{volume}{265}}, \bibinfo{pages}{364} (\bibinfo{year}{1994}).

\bibitem[{\citenamefont{Xie and Dunn}(1994)}]{xie:94}
\bibinfo{author}{\bibfnamefont{X.~S.} \bibnamefont{Xie}} \bibnamefont{and}
  \bibinfo{author}{\bibfnamefont{R.~C.} \bibnamefont{Dunn}},
  \bibinfo{journal}{Science} \textbf{\bibinfo{volume}{265}},
  \bibinfo{pages}{361} (\bibinfo{year}{1994}).

\bibitem[{\citenamefont{Kawata}(1994)}]{Kawata1994}
\bibinfo{author}{\bibfnamefont{S.}~\bibnamefont{Kawata }} \bibnamefont{and}
\bibinfo{author}{\bibfnamefont{Y.}~\bibnamefont{Inouye}},
  \bibinfo{journal}{Opt. Lett.} \textbf{\bibinfo{volume}{19}},
  \bibinfo{pages}{159} (\bibinfo{year}{1994}).

\bibitem[{\citenamefont{Zenhausern et~al.}(1994)\citenamefont{Zenhausern,
  Oboyle, and Wickramasinghe}}]{Zenhausern1994}
\bibinfo{author}{\bibfnamefont{F.}~\bibnamefont{Zenhausern}},
  \bibinfo{author}{\bibfnamefont{M.}~\bibnamefont{Oboyle}}, \bibnamefont{and}
  \bibinfo{author}{\bibfnamefont{H.}~\bibnamefont{Wickramasinghe}},
  \bibinfo{journal}{Appl. Phys. Lett.} \textbf{\bibinfo{volume}{65}},
  \bibinfo{pages}{1623} (\bibinfo{year}{1994}).

\bibitem[{\citenamefont{Gleyzes et~al.}(1995)\citenamefont{Gleyzes, Boccara,
  and Bachelot}}]{gleyzes:95}
\bibinfo{author}{\bibfnamefont{P.}~\bibnamefont{Gleyzes}},
  \bibinfo{author}{\bibfnamefont{A.~C.} \bibnamefont{Boccara}},
  \bibnamefont{and} \bibinfo{author}{\bibfnamefont{R.}~\bibnamefont{Bachelot}},
  \bibinfo{journal}{Ultramicroscopy} \textbf{\bibinfo{volume}{57}},
  \bibinfo{pages}{318} (\bibinfo{year}{1995}).

\bibitem[{\citenamefont{Hartschuh et~al.}(2003)\citenamefont{Hartschuh,
  S\'{a}nchez, Xie, and Novotny}}]{Hartschuh:03}
\bibinfo{author}{\bibfnamefont{A.}~\bibnamefont{Hartschuh}},
  \bibinfo{author}{\bibfnamefont{E.~J.} \bibnamefont{S\'{a}nchez}},
  \bibinfo{author}{\bibfnamefont{X.~S.} \bibnamefont{Xie}}, \bibnamefont{and}
  \bibinfo{author}{\bibfnamefont{L.}~\bibnamefont{Novotny}},
  \bibinfo{journal}{Phys. Rev. Lett.} \textbf{\bibinfo{volume}{90}},
  \bibinfo{pages}{95503} (\bibinfo{year}{2003}).

\bibitem[{\citenamefont{Sanchez et~al.}(1999)\citenamefont{Sanchez, Novotny,
  and Xie}}]{Sanchez:99}
\bibinfo{author}{\bibfnamefont{E.~J.} \bibnamefont{Sanchez}},
  \bibinfo{author}{\bibfnamefont{L.}~\bibnamefont{Novotny}}, \bibnamefont{and}
  \bibinfo{author}{\bibfnamefont{X.~S.} \bibnamefont{Xie}},
  \bibinfo{journal}{Phys. Rev. Lett.} \textbf{\bibinfo{volume}{82}},
  \bibinfo{pages}{4014} (\bibinfo{year}{1999}).

\bibitem[{\citenamefont{Kalkbrenner et~al.}(2001)\citenamefont{Kalkbrenner,
  Ramstein, Mlynek, and Sandoghdar}}]{Kalkbrenner:01}
\bibinfo{author}{\bibfnamefont{T.}~\bibnamefont{Kalkbrenner}},
  \bibinfo{author}{\bibfnamefont{M.}~\bibnamefont{Ramstein}},
  \bibinfo{author}{\bibfnamefont{J.}~\bibnamefont{Mlynek}}, \bibnamefont{and}
  \bibinfo{author}{\bibfnamefont{V.}~\bibnamefont{Sandoghdar}},
  \bibinfo{journal}{J. Micros.} \textbf{\bibinfo{volume}{202}},
  \bibinfo{pages}{72} (\bibinfo{year}{2001}).

\bibitem[{\citenamefont{Wessel}(1985)}]{wessel:85}
\bibinfo{author}{\bibfnamefont{J.}~\bibnamefont{Wessel}}, \bibinfo{journal}{J.
  Opt. Soc. Am. B} \textbf{\bibinfo{volume}{2}}, \bibinfo{pages}{1538}
  (\bibinfo{year}{1985}).

\bibitem[{\citenamefont{Pohl}(2000)}]{Pohl:00}
\bibinfo{author}{\bibfnamefont{D.}~\bibnamefont{Pohl}},
  \emph{\bibinfo{title}{Near-field Optics: Principles and Applications}}
  (\bibinfo{publisher}{World Scientific Publ.}, \bibinfo{address}{Singapore},
  \bibinfo{year}{2000}), chap. \bibinfo{chapter}{Near-field optics seen as an
  antenna problem}, pp. \bibinfo{pages}{9--21}.

\bibitem[{\citenamefont{Schuck et~al.}(2005)\citenamefont{Schuck, Fromm,
  Sundaramurthy, Kino, and Moerner}}]{Schuck:05}
\bibinfo{author}{\bibfnamefont{P.~J.} \bibnamefont{Schuck}},
  \bibinfo{author}{\bibfnamefont{D.~P.} \bibnamefont{Fromm}},
  \bibinfo{author}{\bibfnamefont{A.}~\bibnamefont{Sundaramurthy}},
  \bibinfo{author}{\bibfnamefont{G.~S.} \bibnamefont{Kino}}, \bibnamefont{and}
  \bibinfo{author}{\bibfnamefont{W.~E.} \bibnamefont{Moerner}},
  \bibinfo{journal}{Phys. Rev. Lett.} \textbf{\bibinfo{volume}{94}},
  \bibinfo{pages}{017402} (\bibinfo{year}{2005}).

\bibitem[{\citenamefont{M{\"u}hlschlegel
  et~al.}(2005)\citenamefont{M{\"u}hlschlegel, Eisler, Martin, Hecht, and
  Pohl}}]{Muhlschlegel2005}
\bibinfo{author}{\bibfnamefont{P.}~\bibnamefont{M{\"u}hlschlegel}},
  \bibinfo{author}{\bibfnamefont{H.-J.} \bibnamefont{Eisler}},
  \bibinfo{author}{\bibfnamefont{O.}~\bibnamefont{Martin}},
  \bibinfo{author}{\bibfnamefont{B.}~\bibnamefont{Hecht}}, \bibnamefont{and}
  \bibinfo{author}{\bibfnamefont{D.}~\bibnamefont{Pohl}},
  \bibinfo{journal}{Science} \textbf{\bibinfo{volume}{308}},
  \bibinfo{pages}{1607} (\bibinfo{year}{2005}).

\bibitem[{\citenamefont{Kreibig and Vollmer}(1995)}]{Kreibig}
\bibinfo{author}{\bibfnamefont{U.}~\bibnamefont{Kreibig}} \bibnamefont{and}
  \bibinfo{author}{\bibfnamefont{M.}~\bibnamefont{Vollmer}},
  \emph{\bibinfo{title}{Optical Properties of Metal Clusters}}
  (\bibinfo{publisher}{Springer Berlin}, \bibinfo{year}{1995}).

\bibitem[{\citenamefont{Thomas et~al.}(2004)\citenamefont{Thomas, Greffet,
  Carminati, and Arias-Gonzalez}}]{Thomas2004}
\bibinfo{author}{\bibfnamefont{M.}~\bibnamefont{Thomas}},
  \bibinfo{author}{\bibfnamefont{J.-J.} \bibnamefont{Greffet}},
  \bibinfo{author}{\bibfnamefont{R.}~\bibnamefont{Carminati}},
  \bibnamefont{and}
  \bibinfo{author}{\bibfnamefont{J.}~\bibnamefont{Arias-Gonzalez}},
  \bibinfo{journal}{Appl. Phys. Lett.} \textbf{\bibinfo{volume}{85}},
  \bibinfo{pages}{3863} (\bibinfo{year}{2004}).

\bibitem[{\citenamefont{Bohren and Huffman}(1998)}]{Bohren1998}
\bibinfo{author}{\bibfnamefont{C.}~\bibnamefont{Bohren}} \bibnamefont{and}
  \bibinfo{author}{\bibfnamefont{D.}~\bibnamefont{Huffman}},
  \emph{\bibinfo{title}{Absorption and scattering of light by small particles}}
  (\bibinfo{publisher}{John Wiley \& Sons, Inc.}, \bibinfo{year}{1998}).

\bibitem[{\citenamefont{Tai}(1971)}]{Tai1971}
\bibinfo{author}{\bibfnamefont{C.-T.} \bibnamefont{Tai}},
  \emph{\bibinfo{title}{Dyadic Greens functions in electromagnetic theory}}
  (\bibinfo{publisher}{Intext educational publishers Scranton-San
  Francisco-Toronto-London}, \bibinfo{year}{1971}).

\bibitem[{\citenamefont{Zvyagin and Goto}(1998)}]{Zvyagin1998}
\bibinfo{author}{\bibfnamefont{A.}~\bibnamefont{Zvyagin}} \bibnamefont{and}
  \bibinfo{author}{\bibfnamefont{K.}~\bibnamefont{Goto}}, \bibinfo{journal}{J.
  Opt. Soc. Am. A} \textbf{\bibinfo{volume}{15}}, \bibinfo{pages}{3003}
  (\bibinfo{year}{1998}).

\bibitem[{\citenamefont{Quinten et~al.}(1999)\citenamefont{Quinten, Pack, and
  Wannemacher}}]{Quinten1999}
\bibinfo{author}{\bibfnamefont{M.}~\bibnamefont{Quinten}},
  \bibinfo{author}{\bibfnamefont{A.}~\bibnamefont{Pack}}, \bibnamefont{and}
  \bibinfo{author}{\bibfnamefont{R.}~\bibnamefont{Wannemacher}},
  \bibinfo{journal}{Appl. Phys. B} \textbf{\bibinfo{volume}{68}},
  \bibinfo{pages}{87} (\bibinfo{year}{1999}).

\bibitem[{\citenamefont{Lock}(1995)}]{Lock1995}
\bibinfo{author}{\bibfnamefont{J.}~\bibnamefont{Lock}}, \bibinfo{journal}{J.
  Opt. Soc. Am. A} \textbf{\bibinfo{volume}{12}}, \bibinfo{pages}{929}
  (\bibinfo{year}{1995}).

\bibitem[{\citenamefont{Kelley et~al.}(2003)\citenamefont{Kelley, Coronado,
  Zhao, and Schatz}}]{Kelley2003}
\bibinfo{author}{\bibfnamefont{K.}~\bibnamefont{Kelley}},
  \bibinfo{author}{\bibfnamefont{E.}~\bibnamefont{Coronado}},
  \bibinfo{author}{\bibfnamefont{L.}~\bibnamefont{Zhao}}, \bibnamefont{and}
  \bibinfo{author}{\bibfnamefont{G.}~\bibnamefont{Schatz}},
  \bibinfo{journal}{J. Phys. Chem. B} \textbf{\bibinfo{volume}{107}},
  \bibinfo{pages}{668} (\bibinfo{year}{2003}).

\bibitem[{\citenamefont{Richter et~al.}(1999)\citenamefont{Richter, Jordan,
  Cavanagh, Bryant, Liu, Stranick, Keating, and Natan}}]{Richter1999}
\bibinfo{author}{\bibfnamefont{L.~J.} \bibnamefont{Richter}},
  \bibinfo{author}{\bibfnamefont{C.~E.} \bibnamefont{Jordan}},
  \bibinfo{author}{\bibfnamefont{R.~R.} \bibnamefont{Cavanagh}},
  \bibinfo{author}{\bibfnamefont{G.~W.} \bibnamefont{Bryant}},
  \bibinfo{author}{\bibfnamefont{A.}~\bibnamefont{Liu}},
  \bibinfo{author}{\bibfnamefont{S.~J.} \bibnamefont{Stranick}},
  \bibinfo{author}{\bibfnamefont{C.~D.} \bibnamefont{Keating}},
  \bibnamefont{and} \bibinfo{author}{\bibfnamefont{M.~J.} \bibnamefont{Natan}},
  \bibinfo{journal}{J. Opt. Soc. Am. A} \textbf{\bibinfo{volume}{16}},
  \bibinfo{pages}{1936} (\bibinfo{year}{1999}).

\bibitem[{\citenamefont{Chew et~al.}(1979)\citenamefont{Chew, Wang, and
  Kerker}}]{Chew1979}
\bibinfo{author}{\bibfnamefont{H.}~\bibnamefont{Chew}},
  \bibinfo{author}{\bibfnamefont{D.-S.} \bibnamefont{Wang}}, \bibnamefont{and}
  \bibinfo{author}{\bibfnamefont{M.}~\bibnamefont{Kerker}},
  \bibinfo{journal}{Appl. Opt.} \textbf{\bibinfo{volume}{18}},
  \bibinfo{pages}{2679} (\bibinfo{year}{1979}).

\bibitem[{\citenamefont{Lakowicz}(2005)}]{Lakowicz2005}
\bibinfo{author}{\bibfnamefont{J.~R.} \bibnamefont{Lakowicz}},
  \bibinfo{journal}{Anal. Biochem.} \textbf{\bibinfo{volume}{337}},
  \bibinfo{pages}{171} (\bibinfo{year}{2005}).

\bibitem[{\citenamefont{Bian et~al.}(1995)\citenamefont{Bian, Dunn, Xie, and
  Leung}}]{Bian1995}
\bibinfo{author}{\bibfnamefont{R.~X.} \bibnamefont{Bian}},
  \bibinfo{author}{\bibfnamefont{R.~C.} \bibnamefont{Dunn}},
  \bibinfo{author}{\bibfnamefont{X.~S.} \bibnamefont{Xie}}, \bibnamefont{and}
  \bibinfo{author}{\bibfnamefont{P.~T.} \bibnamefont{Leung}},
  \bibinfo{journal}{Phys. Rev. Lett.} \textbf{\bibinfo{volume}{75}},
  \bibinfo{pages}{4772} (\bibinfo{year}{1995}).

\bibitem[{\citenamefont{Sullivan and Hall}(1997)}]{Sullivan1997}
\bibinfo{author}{\bibfnamefont{K.}~\bibnamefont{Sullivan}} \bibnamefont{and}
  \bibinfo{author}{\bibfnamefont{D.}~\bibnamefont{Hall}},
  \bibinfo{journal}{J. Opt. Soc. Am. B} \textbf{\bibinfo{volume}{14}},
  \bibinfo{pages}{1149} (\bibinfo{year}{1997}).

\bibitem[{\citenamefont{Metiu}(1984{\natexlab{b}})}]{Metiu1984}
\bibinfo{author}{\bibfnamefont{H.}~\bibnamefont{Metiu}},
  \bibinfo{journal}{Prog. Surf. Sci.} \textbf{\bibinfo{volume}{17}},
  \bibinfo{pages}{153} (\bibinfo{year}{1984}{\natexlab{b}}).

\bibitem[{\citenamefont{Klimov et~al.}(1996)\citenamefont{Klimov, Ducloy, and
  Letokov}}]{Klimov1996}
\bibinfo{author}{\bibfnamefont{V.}~\bibnamefont{Klimov}},
  \bibinfo{author}{\bibfnamefont{M.}~\bibnamefont{Ducloy}}, \bibnamefont{and}
  \bibinfo{author}{\bibfnamefont{V.}~\bibnamefont{Letokov}},
  \bibinfo{journal}{J. Mod. Phys.} \textbf{\bibinfo{volume}{43}},
  \bibinfo{pages}{2251} (\bibinfo{year}{1996}).

\bibitem[{\citenamefont{Chew}(1987)}]{Chew1987}
\bibinfo{author}{\bibfnamefont{H.}~\bibnamefont{Chew}},
  \bibinfo{journal}{J. Chem. Phys.} \textbf{\bibinfo{volume}{87}},
  \bibinfo{pages}{1355} (\bibinfo{year}{1987}).

\bibitem[{\citenamefont{Das and Puri}(2002)}]{Das2002}
\bibinfo{author}{\bibfnamefont{P.}~\bibnamefont{Das}} \bibnamefont{and}
  \bibinfo{author}{\bibfnamefont{A.}~\bibnamefont{Puri}},
  \bibinfo{journal}{Phys. Rev. B} \textbf{\bibinfo{volume}{65}},
  \bibinfo{pages}{155416} (\bibinfo{year}{2002}).

\bibitem[{\citenamefont{Agio et~al.}(2007)\citenamefont{Agio, Mori, Kaminski,
  Rogobete, K\"uhn, Callegari, Nellen, Robin, Ekinci, Sennhauser
  et~al.}}]{Agio2007}
\bibinfo{author}{\bibfnamefont{M.}~\bibnamefont{Agio}},
  \bibinfo{author}{\bibfnamefont{G.}~\bibnamefont{Mori}},
  \bibinfo{author}{\bibfnamefont{F.}~\bibnamefont{Kaminski}},
  \bibinfo{author}{\bibfnamefont{L.}~\bibnamefont{Rogobete}},
  \bibinfo{author}{\bibfnamefont{S.}~\bibnamefont{K\"uhn}},
  \bibinfo{author}{\bibfnamefont{V.}~\bibnamefont{Callegari}},
  \bibinfo{author}{\bibfnamefont{P.~M.} \bibnamefont{Nellen}},
  \bibinfo{author}{\bibfnamefont{F.}~\bibnamefont{Robin}},
  \bibinfo{author}{\bibfnamefont{Y.}~\bibnamefont{Ekinci}},
  \bibinfo{author}{\bibfnamefont{U.}~\bibnamefont{Sennhauser}},
  \bibnamefont{et~al.} \bibinfo{journal}{Proc. SPIE}, submitted.

\bibitem[{\citenamefont{Caticha}(2000)}]{Caticha2000}
\bibinfo{author}{\bibfnamefont{A.}~\bibnamefont{Caticha}},
  \bibinfo{journal}{Phys. Rev. B} \textbf{\bibinfo{volume}{62}},
  \bibinfo{pages}{3639} (\bibinfo{year}{2000}).

\bibitem[{\citenamefont{Carminati and Nieto-Vesperinas}(1998)}]{Carminati1998}
\bibinfo{author}{\bibfnamefont{R.}~\bibnamefont{Carminati}} \bibnamefont{and}
  \bibinfo{author}{\bibfnamefont{M.}~\bibnamefont{Nieto-Vesperinas}},
  \bibinfo{journal}{J. Opt. Soc. Am. A} \textbf{\bibinfo{volume}{15}},
  \bibinfo{pages}{706} (\bibinfo{year}{1998}).

\bibitem[{\citenamefont{Betzig et~al.}(1992)\citenamefont{Betzig, Finn, and
  Weiner}}]{Betzig1992}
\bibinfo{author}{\bibfnamefont{E.}~\bibnamefont{Betzig}},
  \bibinfo{author}{\bibfnamefont{P.}~\bibnamefont{Finn}}, \bibnamefont{and}
  \bibinfo{author}{\bibfnamefont{J.}~\bibnamefont{Weiner}},
  \bibinfo{journal}{Appl. Phys. Lett.} \textbf{\bibinfo{volume}{60}},
  \bibinfo{pages}{2484} (\bibinfo{year}{1992}).

\bibitem[{\citenamefont{Karrai and Tiemann}(2000)}]{Karrai2000}
\bibinfo{author}{\bibfnamefont{K.}~\bibnamefont{Karrai}} \bibnamefont{and}
  \bibinfo{author}{\bibfnamefont{I.}~\bibnamefont{Tiemann}},
  \bibinfo{journal}{Phys. Rev. B} \textbf{\bibinfo{volume}{62}},
  \bibinfo{pages}{13174} (\bibinfo{year}{2000}).

\bibitem[{\citenamefont{Pfab et~al.}(2004)\citenamefont{Pfab, Zimmermann,
  Hettich, Gerhardt, Renn, and Sandoghdar}}]{Pfab:04}
\bibinfo{author}{\bibfnamefont{R.~J.} \bibnamefont{Pfab}},
  \bibinfo{author}{\bibfnamefont{J.}~\bibnamefont{Zimmermann}},
  \bibinfo{author}{\bibfnamefont{C.}~\bibnamefont{Hettich}},
  \bibinfo{author}{\bibfnamefont{I.}~\bibnamefont{Gerhardt}},
  \bibinfo{author}{\bibfnamefont{A.}~\bibnamefont{Renn}}, \bibnamefont{and}
  \bibinfo{author}{\bibfnamefont{V.}~\bibnamefont{Sandoghdar}},
  \bibinfo{journal}{Chem. Phys. Lett.} \textbf{\bibinfo{volume}{387}},
  \bibinfo{pages}{490} (\bibinfo{year}{2004}).

\bibitem[{\citenamefont{Buchler et~al.}(2005)\citenamefont{Buchler,
  Kalkbrenner, Hettich, and Sandoghdar}}]{Buchler:05}
\bibinfo{author}{\bibfnamefont{B.~C.} \bibnamefont{Buchler}},
  \bibinfo{author}{\bibfnamefont{T.}~\bibnamefont{Kalkbrenner}},
  \bibinfo{author}{\bibfnamefont{C.}~\bibnamefont{Hettich}}, \bibnamefont{and}
  \bibinfo{author}{\bibfnamefont{V.}~\bibnamefont{Sandoghdar}},
  \bibinfo{journal}{Phys. Rev. Lett.} \textbf{\bibinfo{volume}{95}},
  \bibinfo{pages}{063003} (\bibinfo{year}{2005}).

\bibitem[{\citenamefont{Kummer et~al.}(1997)\citenamefont{Kummer, Kulzer,
  Kettner, Basche, Tietz, Glowatz, and Kryschi}}]{Kummer1997}
\bibinfo{author}{\bibfnamefont{S.}~\bibnamefont{Kummer}},
  \bibinfo{author}{\bibfnamefont{F.}~\bibnamefont{Kulzer}},
  \bibinfo{author}{\bibfnamefont{R.}~\bibnamefont{Kettner}},
  \bibinfo{author}{\bibfnamefont{T.}~\bibnamefont{Basche}},
  \bibinfo{author}{\bibfnamefont{C.}~\bibnamefont{Tietz}},
  \bibinfo{author}{\bibfnamefont{C.}~\bibnamefont{Glowatz}}, \bibnamefont{and}
  \bibinfo{author}{\bibfnamefont{C.}~\bibnamefont{Kryschi}},
  \bibinfo{journal}{J. Chem. Phys.} \textbf{\bibinfo{volume}{107}},
  \bibinfo{pages}{7673} (\bibinfo{year}{1997}).

\bibitem[{\citenamefont{Becker et~al.}(2004)\citenamefont{Becker, Bergmann,
  Hink, K{\"o}nig, Benndorf, and Biskup}}]{Becker2004}
\bibinfo{author}{\bibfnamefont{W.}~\bibnamefont{Becker}},
  \bibinfo{author}{\bibfnamefont{A.}~\bibnamefont{Bergmann}},
  \bibinfo{author}{\bibfnamefont{M.}~\bibnamefont{Hink}},
  \bibinfo{author}{\bibfnamefont{K.}~\bibnamefont{K{\"o}nig}},
  \bibinfo{author}{\bibfnamefont{K.}~\bibnamefont{Benndorf}}, \bibnamefont{and}
  \bibinfo{author}{\bibfnamefont{C.}~\bibnamefont{Biskup}},
  \bibinfo{journal}{Microscopy Research and Technique}
  \textbf{\bibinfo{volume}{63}}, \bibinfo{pages}{58} (\bibinfo{year}{2004}).

\bibitem[{\citenamefont{Bohmer and Enderlein}(2004)}]{Bohmer2004}
\bibinfo{author}{\bibfnamefont{M.}~\bibnamefont{Bohmer}} \bibnamefont{and}
  \bibinfo{author}{\bibfnamefont{J.}~\bibnamefont{Enderlein}},
  \bibinfo{journal}{J. Opt. Soc. Am. B} \textbf{\bibinfo{volume}{20}},
  \bibinfo{pages}{554} (\bibinfo{year}{2004}).

\bibitem[{\citenamefont{Haroche}(1992)}]{Haroche-Houches}
\bibinfo{author}{\bibfnamefont{S.}~\bibnamefont{Haroche}},
  in \emph{\bibinfo{title}{Fundamental systems in quantum optics}}
  (\bibinfo{publisher}{North-Holland}, \bibinfo{address}{Amsterdam},
  \bibinfo{year}{1992}), pp. \bibinfo{pages}{767--940}.

\bibitem[{\citenamefont{Trabesinger et~al.}(2003)\citenamefont{Trabesinger,
  Kramer, Kreiter, Hecht, and Wild}}]{Trabesinger:03b}
\bibinfo{author}{\bibfnamefont{W.}~\bibnamefont{Trabesinger}},
  \bibinfo{author}{\bibfnamefont{A.}~\bibnamefont{Kramer}},
  \bibinfo{author}{\bibfnamefont{M.}~\bibnamefont{Kreiter}},
  \bibinfo{author}{\bibfnamefont{B.}~\bibnamefont{Hecht}}, \bibnamefont{and}
  \bibinfo{author}{\bibfnamefont{U.}~\bibnamefont{Wild}}, \bibinfo{journal}{J.
  Microscopy} \textbf{\bibinfo{volume}{209}}, \bibinfo{pages}{249}
  (\bibinfo{year}{2003}).

\bibitem[{\citenamefont{Trabesinger et~al.}(2002)\citenamefont{Trabesinger,
  Kramer, Kreiter, Hecht, and Wild}}]{Trabesinger:02}
\bibinfo{author}{\bibfnamefont{W.}~\bibnamefont{Trabesinger}},
  \bibinfo{author}{\bibfnamefont{A.}~\bibnamefont{Kramer}},
  \bibinfo{author}{\bibfnamefont{M.}~\bibnamefont{Kreiter}},
  \bibinfo{author}{\bibfnamefont{B.}~\bibnamefont{Hecht}}, \bibnamefont{and}
  \bibinfo{author}{\bibfnamefont{U.~P.} \bibnamefont{Wild}},
  \bibinfo{journal}{Appl. Phys. Lett.} \textbf{\bibinfo{volume}{81}},
  \bibinfo{pages}{2118} (\bibinfo{year}{2002}).

\bibitem[{\citenamefont{Protasenko and Gallagher}(2004)}]{Protasenko2004}
\bibinfo{author}{\bibfnamefont{V.}~\bibnamefont{Protasenko}} \bibnamefont{and}
  \bibinfo{author}{\bibfnamefont{A.}~\bibnamefont{Gallagher}},
  \bibinfo{journal}{Nano Lett.} \textbf{\bibinfo{volume}{4}},
  \bibinfo{pages}{1329} (\bibinfo{year}{2004}).

\bibitem[{\citenamefont{H'dhili et~al.}(2002)\citenamefont{H'dhili, Bachelot,
  Rumyantseva, and Lerondel}}]{H'dhili2002}
\bibinfo{author}{\bibfnamefont{F.}~\bibnamefont{H'dhili}},
  \bibinfo{author}{\bibfnamefont{R.}~\bibnamefont{Bachelot}},
  \bibinfo{author}{\bibfnamefont{A.}~\bibnamefont{Rumyantseva}},
  \bibnamefont{and} \bibinfo{author}{\bibfnamefont{G.}~\bibnamefont{Lerondel}},
  \bibinfo{journal}{J. Microscopy} \textbf{\bibinfo{volume}{209}},
  \bibinfo{pages}{214} (\bibinfo{year}{2002}).

\bibitem[{\citenamefont{Martin et~al.}(2001)\citenamefont{Martin, Hamann, and
  Wickramasinghe}}]{Martin2001}
\bibinfo{author}{\bibfnamefont{Y.~C.} \bibnamefont{Martin}},
  \bibinfo{author}{\bibfnamefont{H.~F.} \bibnamefont{Hamann}},
  \bibnamefont{and} \bibinfo{author}{\bibfnamefont{H.~K.}
  \bibnamefont{Wickramasinghe}}, \bibinfo{journal}{J. Appl. Phys.}
  \textbf{\bibinfo{volume}{89}}, \bibinfo{pages}{5774} (\bibinfo{year}{2001}).

\bibitem[{\citenamefont{Patane et~al.}(2004)\citenamefont{Patane, Gucciardi,
  Labardi, and Allegrini}}]{Patane2004}
\bibinfo{author}{\bibfnamefont{S.}~\bibnamefont{Patane}},
  \bibinfo{author}{\bibfnamefont{P.}~\bibnamefont{Gucciardi}},
  \bibinfo{author}{\bibfnamefont{M.}~\bibnamefont{Labardi}}, \bibnamefont{and}
  \bibinfo{author}{\bibfnamefont{M.}~\bibnamefont{Allegrini}},
  \bibinfo{journal}{Revista del nuovo cimento} \textbf{\bibinfo{volume}{27}},
  \bibinfo{pages}{1} (\bibinfo{year}{2004}).

\bibitem[{\citenamefont{Chance et~al.}(1978)\citenamefont{Chance, Prock, and
  Silbey}}]{Chance1978}
\bibinfo{author}{\bibfnamefont{R.}~\bibnamefont{Chance}},
  \bibinfo{author}{\bibfnamefont{A.}~\bibnamefont{Prock}}, \bibnamefont{and}
  \bibinfo{author}{\bibfnamefont{R.}~\bibnamefont{Silbey}},
  \bibinfo{journal}{Adv. Chem. Phys.} \textbf{\bibinfo{volume}{37}},
  \bibinfo{pages}{1} (\bibinfo{year}{1978}).

\bibitem[{\citenamefont{Ermushev et~al.}(1993)\citenamefont{Ermushev,
  Mchedlishvili, Oleinikov, and Petukhov}}]{Ermushev1993}
\bibinfo{author}{\bibfnamefont{A.}~\bibnamefont{Ermushev}},
  \bibinfo{author}{\bibfnamefont{B.}~\bibnamefont{Mchedlishvili}},
  \bibinfo{author}{\bibfnamefont{V.}~\bibnamefont{Oleinikov}},
  \bibnamefont{and} \bibinfo{author}{\bibfnamefont{A.}~\bibnamefont{Petukhov}},
  \bibinfo{journal}{Quant. Electron.} \textbf{\bibinfo{volume}{23}},
  \bibinfo{pages}{435} (\bibinfo{year}{1993}).

\bibitem[{\citenamefont{Lide}(1995)}]{Lide1995}
\bibinfo{author}{\bibfnamefont{S.~R.} \bibnamefont{Lide}},
  \emph{\bibinfo{title}{Handbook of {C}hemistry and {P}hysics, 75th edn.}}
  (\bibinfo{publisher}{CRC Press}, \bibinfo{year}{1995}).

\bibitem[{\citenamefont{Klimov et~al.}(2001)\citenamefont{Klimov, Ducloy, and
  Letokov}}]{Klimov2001}
\bibinfo{author}{\bibfnamefont{V.}~\bibnamefont{Klimov}},
  \bibinfo{author}{\bibfnamefont{M.}~\bibnamefont{Ducloy}}, \bibnamefont{and}
  \bibinfo{author}{\bibfnamefont{V.}~\bibnamefont{Letokov}},
  \bibinfo{journal}{Quant. Electron.} \textbf{\bibinfo{volume}{31}},
  \bibinfo{pages}{569} (\bibinfo{year}{2001}).

\bibitem[{\citenamefont{Azoulay et~al.}(2000)\citenamefont{Azoulay, Debarre,
  Richard, and Tchenio}}]{Azoulay2000}
\bibinfo{author}{\bibfnamefont{J.}~\bibnamefont{Azoulay}},
  \bibinfo{author}{\bibfnamefont{A.}~\bibnamefont{Debarre}},
  \bibinfo{author}{\bibfnamefont{A.}~\bibnamefont{Richard}}, \bibnamefont{and}
  \bibinfo{author}{\bibfnamefont{P.}~\bibnamefont{Tchenio}},
  \bibinfo{journal}{Europhys. Lett.} \textbf{\bibinfo{volume}{51}},
  \bibinfo{pages}{374} (\bibinfo{year}{2000}).

\bibitem[{\citenamefont{Inouye and Kawata}(1994)}]{inouye:94}
\bibinfo{author}{\bibfnamefont{Y.}~\bibnamefont{Inouye}} \bibnamefont{and}
  \bibinfo{author}{\bibfnamefont{S.}~\bibnamefont{Kawata}},
  \bibinfo{journal}{Opt. Lett.} \textbf{\bibinfo{volume}{19}},
  \bibinfo{pages}{159} (\bibinfo{year}{1994}).

\bibitem[{\citenamefont{Kramer et~al.}(2002)\citenamefont{Kramer, Trabesinger,
  Hecht, and Wild}}]{Kramer2002}
\bibinfo{author}{\bibfnamefont{A.}~\bibnamefont{Kramer}},
  \bibinfo{author}{\bibfnamefont{W.}~\bibnamefont{Trabesinger}},
  \bibinfo{author}{\bibfnamefont{B.}~\bibnamefont{Hecht}}, \bibnamefont{and}
  \bibinfo{author}{\bibfnamefont{U.}~\bibnamefont{Wild}},
  \bibinfo{journal}{Appl. Phys. Lett.} \textbf{\bibinfo{volume}{80}},
  \bibinfo{pages}{1652} (\bibinfo{year}{2002}).

\bibitem[{\citenamefont{Kalkbrenner et~al.}(2005)\citenamefont{Kalkbrenner,
  H\aa{}kanson, Sch\"{a}dle, Burger, Henkel, and Sandoghdar}}]{Kalkbrenner:05}
\bibinfo{author}{\bibfnamefont{T.}~\bibnamefont{Kalkbrenner}},
  \bibinfo{author}{\bibfnamefont{U.}~\bibnamefont{H\aa{}kanson}},
  \bibinfo{author}{\bibfnamefont{A.}~\bibnamefont{Sch\"{a}dle}},
  \bibinfo{author}{\bibfnamefont{S.}~\bibnamefont{Burger}},
  \bibinfo{author}{\bibfnamefont{C.}~\bibnamefont{Henkel}}, \bibnamefont{and}
  \bibinfo{author}{\bibfnamefont{V.}~\bibnamefont{Sandoghdar}},
  \bibinfo{journal}{Phys. Rev. Lett.} \textbf{\bibinfo{volume}{95}},
  \bibinfo{pages}{200801} (\bibinfo{year}{2005}).

\bibitem[{\citenamefont{Frey et~al.}(2004)\citenamefont{Frey, Witt, Felderer,
  and Guckenberger}}]{Frey2004}
\bibinfo{author}{\bibfnamefont{H.}~\bibnamefont{Frey}},
  \bibinfo{author}{\bibfnamefont{S.}~\bibnamefont{Witt}},
  \bibinfo{author}{\bibfnamefont{K.}~\bibnamefont{Felderer}}, \bibnamefont{and}
  \bibinfo{author}{\bibfnamefont{R.}~\bibnamefont{Guckenberger}},
  \bibinfo{journal}{Phys. Rev. Lett.} \textbf{\bibinfo{volume}{93}},
  \bibinfo{pages}{200801} (\bibinfo{year}{2004}).

\bibitem[{\citenamefont{K{\"u}hn and Sandoghdar}(2007)}]{Kuehn2007}
\bibinfo{author}{\bibfnamefont{S.}~\bibnamefont{K{\"u}hn}} \bibnamefont{and}
  \bibinfo{author}{\bibfnamefont{V.}~\bibnamefont{Sandoghdar}}
  (\bibinfo{year}{2007}), \bibinfo{note}{in preparation}.

\bibitem[{\citenamefont{Weeber et~al.}(1999)\citenamefont{Weeber, Girard,
  Krenn, Dereux, and Goudonnet}}]{Weeber1999}
\bibinfo{author}{\bibfnamefont{J.-C.} \bibnamefont{Weeber}},
  \bibinfo{author}{\bibfnamefont{C.}~\bibnamefont{Girard}},
  \bibinfo{author}{\bibfnamefont{J.~R.} \bibnamefont{Krenn}},
  \bibinfo{author}{\bibfnamefont{A.}~\bibnamefont{Dereux}}, \bibnamefont{and}
  \bibinfo{author}{\bibfnamefont{J.-P.} \bibnamefont{Goudonnet}},
  \bibinfo{journal}{J. Appl. Phys.} \textbf{\bibinfo{volume}{86}},
  \bibinfo{pages}{2576} (\bibinfo{year}{1999}).

\bibitem[{\citenamefont{Anger et~al.}(2006{\natexlab{a}})\citenamefont{Anger,
  Bharadwaj, and Novotny}}]{Anger2006}
\bibinfo{author}{\bibfnamefont{P.}~\bibnamefont{Anger}},
  \bibinfo{author}{\bibfnamefont{P.}~\bibnamefont{Bharadwaj}},
  \bibnamefont{and} \bibinfo{author}{\bibfnamefont{L.}~\bibnamefont{Novotny}},
  \bibinfo{journal}{Phys. Rev. Lett.} \textbf{\bibinfo{volume}{96}},
  \bibinfo{pages}{113002} (\bibinfo{year}{2006}{\natexlab{a}}).

\bibitem[{\citenamefont{Christ et~al.}(2001)\citenamefont{Christ, Kulzer,
  Bordat, and Basche}}]{Christ2001}
\bibinfo{author}{\bibfnamefont{T.}~\bibnamefont{Christ}},
  \bibinfo{author}{\bibfnamefont{F.}~\bibnamefont{Kulzer}},
  \bibinfo{author}{\bibfnamefont{P.}~\bibnamefont{Bordat}}, \bibnamefont{and}
  \bibinfo{author}{\bibfnamefont{T.}~\bibnamefont{Basche}},
  \bibinfo{journal}{Ang. Chem. Int. Ed.} \textbf{\bibinfo{volume}{40}},
  \bibinfo{pages}{4192} (\bibinfo{year}{2001}).

\bibitem[{\citenamefont{Renn et~al.}(2006)\citenamefont{Renn, Seelig, and
  Sandoghdar}}]{Renn2006}
\bibinfo{author}{\bibfnamefont{A.}~\bibnamefont{Renn}},
  \bibinfo{author}{\bibfnamefont{J.}~\bibnamefont{Seelig}}, \bibnamefont{and}
  \bibinfo{author}{\bibfnamefont{V.}~\bibnamefont{Sandoghdar}},
  \bibinfo{journal}{Mol. Phys.} \textbf{\bibinfo{volume}{104}},
  \bibinfo{pages}{409} (\bibinfo{year}{2006}).

\bibitem[{\citenamefont{Anger et~al.}(2006{\natexlab{b}})\citenamefont{Anger,
  Bharadwaj, and Novotny}}]{Anger:06}
\bibinfo{author}{\bibfnamefont{P.}~\bibnamefont{Anger}},
  \bibinfo{author}{\bibfnamefont{P.}~\bibnamefont{Bharadwaj}},
  \bibnamefont{and} \bibinfo{author}{\bibfnamefont{L.}~\bibnamefont{Novotny}},
  \bibinfo{journal}{Phys. Rev. Lett.} \textbf{\bibinfo{volume}{96}},
  \bibinfo{pages}{113002} (\bibinfo{year}{2006}{\natexlab{b}}).

\bibitem[{\citenamefont{Alda et~al.}(2005)\citenamefont{Alda, Rico-Garcia,
  Lopez-Alonso, and Boreman}}]{Alda2005}
\bibinfo{author}{\bibfnamefont{J.}~\bibnamefont{Alda}},
  \bibinfo{author}{\bibfnamefont{J.}~\bibnamefont{Rico-Garcia}},
  \bibinfo{author}{\bibfnamefont{J.}~\bibnamefont{Lopez-Alonso}},
  \bibnamefont{and} \bibinfo{author}{\bibfnamefont{G.}~\bibnamefont{Boreman}},
  \bibinfo{journal}{Egypt. J. Solids} \textbf{\bibinfo{volume}{28}},
  \bibinfo{pages}{1} (\bibinfo{year}{2005}).

\bibitem[{\citenamefont{Crozier et~al.}(2003)\citenamefont{Crozier,
  Sundaramurthy, Kino, and Quate}}]{Crozier2003}
\bibinfo{author}{\bibfnamefont{K.}~\bibnamefont{Crozier}},
  \bibinfo{author}{\bibfnamefont{A.}~\bibnamefont{Sundaramurthy}},
  \bibinfo{author}{\bibfnamefont{G.}~\bibnamefont{Kino}}, \bibnamefont{and}
  \bibinfo{author}{\bibfnamefont{C.}~\bibnamefont{Quate}},
  \bibinfo{journal}{J. Appl. Phys.} \textbf{\bibinfo{volume}{94}},
  \bibinfo{pages}{4632} (\bibinfo{year}{2003}).

\bibitem[{\citenamefont{Greffet}(2005)}]{Greffet2005}
\bibinfo{author}{\bibfnamefont{J.}~\bibnamefont{Greffet}},
  \bibinfo{journal}{Science} \textbf{\bibinfo{volume}{308}},
  \bibinfo{pages}{1561} (\bibinfo{year}{2005}).

\bibitem[{\citenamefont{Joulain et~al.}(2003)\citenamefont{Joulain, Carminati,
  Mulet, and Greffet}}]{Joulain2003}
\bibinfo{author}{\bibfnamefont{K.}~\bibnamefont{Joulain}},
  \bibinfo{author}{\bibfnamefont{R.}~\bibnamefont{Carminati}},
  \bibinfo{author}{\bibfnamefont{J.-P.} \bibnamefont{Mulet}}, \bibnamefont{and}
  \bibinfo{author}{\bibfnamefont{J.-J.} \bibnamefont{Greffet}},
  \bibinfo{journal}{Phys. Rev. B} \textbf{\bibinfo{volume}{68}},
  \bibinfo{pages}{245405} (\bibinfo{year}{2003}).

\bibitem[{\citenamefont{K{\"a}ll et~al.}(2005)\citenamefont{K{\"a}ll, Xu, and
  Johansson}}]{Kaell2005}
\bibinfo{author}{\bibfnamefont{M.}~\bibnamefont{K{\"a}ll}},
  \bibinfo{author}{\bibfnamefont{H.}~\bibnamefont{Xu}}, \bibnamefont{and}
  \bibinfo{author}{\bibfnamefont{P.}~\bibnamefont{Johansson}},
  \bibinfo{journal}{J. Raman Spect.}
  \textbf{\bibinfo{volume}{36}}, \bibinfo{pages}{510} (\bibinfo{year}{2005}).

\bibitem[{\citenamefont{Calander and Willander}(2002)}]{Calander2002}
\bibinfo{author}{\bibfnamefont{N.}~\bibnamefont{Calander}} \bibnamefont{and}
  \bibinfo{author}{\bibfnamefont{M.}~\bibnamefont{Willander}},
  \bibinfo{journal}{J. Appl. Phys.} \textbf{\bibinfo{volume}{92}},
  \bibinfo{pages}{4878} (\bibinfo{year}{2002}).

\bibitem[{\citenamefont{Li et~al.}(2007)\citenamefont{Li, Salandrino, and
  Engheta}}]{Li2007}
\bibinfo{author}{\bibfnamefont{J.}~\bibnamefont{Li}},
  \bibinfo{author}{\bibfnamefont{A.}~\bibnamefont{Salandrino}},
  \bibnamefont{and} \bibinfo{author}{\bibfnamefont{N.}~\bibnamefont{Engheta}}
  (\bibinfo{year}{2007}).

\bibitem[{\citenamefont{Hofmann et~al.}(2007)\citenamefont{Hofmann, Kosako, and
  Kadoya}}]{Hofmann2007}
\bibinfo{author}{\bibfnamefont{H.~F.} \bibnamefont{Hofmann}},
  \bibinfo{author}{\bibfnamefont{T.}~\bibnamefont{Kosako}}, \bibnamefont{and}
  \bibinfo{author}{\bibfnamefont{Y.}~\bibnamefont{Kadoya}},
  \bibinfo{journal}{New J. Phys.} \textbf{\bibinfo{volume}{9}},
  \bibinfo{pages}{217} (\bibinfo{year}{2007}).

\bibitem[{\citenamefont{Gersen et~al.}(2000)\citenamefont{Gersen,
  Garcia-Parajo, Novotny, Veerman, Kuipers, and van Hulst}}]{Gersen2000}
\bibinfo{author}{\bibfnamefont{H.}~\bibnamefont{Gersen}},
  \bibinfo{author}{\bibfnamefont{M.}~\bibnamefont{Garcia-Parajo}},
  \bibinfo{author}{\bibfnamefont{L.}~\bibnamefont{Novotny}},
  \bibinfo{author}{\bibfnamefont{J.}~\bibnamefont{Veerman}},
  \bibinfo{author}{\bibfnamefont{L.}~\bibnamefont{Kuipers}}, \bibnamefont{and}
  \bibinfo{author}{\bibfnamefont{N.}~\bibnamefont{van Hulst}},
  \bibinfo{journal}{Phys. Rev. Lett.} \textbf{\bibinfo{volume}{85}},
  \bibinfo{pages}{5312} (\bibinfo{year}{2000}).

\bibitem[{\citenamefont{Lieb et~al.}(2004)\citenamefont{Lieb, Zavislan, and
  Novotny}}]{Lieb2004}
\bibinfo{author}{\bibfnamefont{M.}~\bibnamefont{Lieb}},
  \bibinfo{author}{\bibfnamefont{J.}~\bibnamefont{Zavislan}}, \bibnamefont{and}
  \bibinfo{author}{\bibfnamefont{L.}~\bibnamefont{Novotny}},
  \bibinfo{journal}{J. Opt. Soc. Am. B} \textbf{\bibinfo{volume}{21}},
  \bibinfo{pages}{1210} (\bibinfo{year}{2004}).

\bibitem[{\citenamefont{Hellen and Axelrod}(1987)}]{Hellen1987}
\bibinfo{author}{\bibfnamefont{E.}~\bibnamefont{Hellen}} \bibnamefont{and}
  \bibinfo{author}{\bibfnamefont{D.}~\bibnamefont{Axelrod}},
  \bibinfo{journal}{J. Opt. Soc. Am. B} \textbf{\bibinfo{volume}{4}},
  \bibinfo{pages}{337} (\bibinfo{year}{1987}).

\bibitem[{\citenamefont{Arnoldus}(2004)}]{Arnoldus2004}
\bibinfo{author}{\bibfnamefont{H.}~\bibnamefont{Arnoldus}},
  \bibinfo{journal}{J. Opt. Soc. Am. A} \textbf{\bibinfo{volume}{22}},
  \bibinfo{pages}{190} (\bibinfo{year}{2004}).

\bibitem[{\citenamefont{Quabis et~al.}(2000)\citenamefont{Quabis, Dorn,
  Eberler, Gl{\"o}ckl, and Leuchs}}]{Quabis2000}
\bibinfo{author}{\bibfnamefont{S.}~\bibnamefont{Quabis}},
  \bibinfo{author}{\bibfnamefont{R.}~\bibnamefont{Dorn}},
  \bibinfo{author}{\bibfnamefont{M.}~\bibnamefont{Eberler}},
  \bibinfo{author}{\bibfnamefont{O.}~\bibnamefont{Gl{\"o}ckl}},
  \bibnamefont{and} \bibinfo{author}{\bibfnamefont{G.}~\bibnamefont{Leuchs}},
  \bibinfo{journal}{Opt. Comm.} \textbf{\bibinfo{volume}{179}},
  \bibinfo{pages}{1} (\bibinfo{year}{2000}).

\bibitem[{\citenamefont{Kaminski et~al.}(2007)\citenamefont{Kaminski,
  Sandoghdar, and Agio}}]{Kaminski2007}
\bibinfo{author}{\bibfnamefont{F.}~\bibnamefont{Kaminski}},
  \bibinfo{author}{\bibfnamefont{V.}~\bibnamefont{Sandoghdar}},
  \bibnamefont{and} \bibinfo{author}{\bibfnamefont{M.}~\bibnamefont{Agio}},
  \bibinfo{journal}{J. Computational and Theoretical Nanoscience}
  \textbf{\bibinfo{volume}{4}}, \bibinfo{pages}{635} (\bibinfo{year}{2007}).

\bibitem[{\citenamefont{Rogobete
  et~al.}(2007{\natexlab{b}})\citenamefont{Rogobete, Kaminski, A.~Mohammadi,
  and Sandoghdar}}]{Rogobete2007a}
\bibinfo{author}{\bibfnamefont{L.}~\bibnamefont{Rogobete}},
  \bibinfo{author}{\bibfnamefont{F.}~\bibnamefont{Kaminski}},
  \bibinfo{author}{\bibnamefont{A.}~\bibnamefont{Mohammadi}},
  \bibinfo{author}{\bibfnamefont{M.}~\bibnamefont{Agio}},
  \bibnamefont{and}
  \bibinfo{author}{\bibfnamefont{V.}~\bibnamefont{Sandoghdar}}
  (\bibinfo{year}{2007}{\natexlab{b}}), \bibinfo{note}{in preparation}.

\bibitem[{\citenamefont{Aravind et~al.}(1981)\citenamefont{Aravind, Nitzan, and
  Metiu}}]{Aravind1981}
\bibinfo{author}{\bibfnamefont{P.~K.} \bibnamefont{Aravind}},
  \bibinfo{author}{\bibfnamefont{A.}~\bibnamefont{Nitzan}}, \bibnamefont{and}
  \bibinfo{author}{\bibfnamefont{H.}~\bibnamefont{Metiu}},
  \bibinfo{journal}{Surf. Science} \textbf{\bibinfo{volume}{110}},
  \bibinfo{pages}{189} (\bibinfo{year}{1981}).

\bibitem[{\citenamefont{Aizpurua et~al.}(2005)\citenamefont{Aizpurua, Bryant,
  Richter, de~Abajo, Kelley, and Mallouk}}]{Aizpurua2005}
\bibinfo{author}{\bibfnamefont{J.}~\bibnamefont{Aizpurua}},
  \bibinfo{author}{\bibfnamefont{G.~W.} \bibnamefont{Bryant}},
  \bibinfo{author}{\bibfnamefont{L.~J.} \bibnamefont{Richter}},
  \bibinfo{author}{\bibfnamefont{F.~J.~G.} \bibnamefont{de~Abajo}},
  \bibinfo{author}{\bibfnamefont{B.~K.} \bibnamefont{Kelley}},
  \bibnamefont{and} \bibinfo{author}{\bibfnamefont{T.}~\bibnamefont{Mallouk}},
  \bibinfo{journal}{Phys. Rev. B} \textbf{\bibinfo{volume}{71}},
  \bibinfo{eid}{235420} (\bibinfo{year}{2005}).

\end{thebibliography}
\end{document}